\documentstyle[11pt]{article}

\begin{document}
\oddsidemargin .3in
\evensidemargin 0 true pt
\topmargin -.4in

%Abbreviations %
%***************%

\def\ra{{\rightarrow}}
\def\a{{\alpha}}
\def\b{{\beta}}
\def\l{{\lambda}}
\def\eps{{\epsilon}}
\def\T{{\Theta}}
\def\t{{\theta}}
\def\co{{\cal O}}
\def\car{{\cal R}}
\def\caf{{\cal F}}
\def\cs{{\Theta_S}}
\def\pr{{\partial}}
\def\tri{{\triangle}}
\def\na{{\nabla }}
\def\S{{\Sigma}}
\def\s{{\sigma}}
\def\sp{\vspace{.15in}}
\def\hs{\hspace{.25in}}

\newcommand{\be}{\begin{equation}} \newcommand{\ee}{\end{equation}}
\newcommand{\bea}{\begin{eqnarray}}\newcommand{\eea}
{\end{eqnarray}}

%********************************************************************%

\begin{titlepage}
\topmargin= -.2in
\textheight 9.5in

%\begin{flushright}
%{hep-th/0508xxx}
%\end{flushright}
%\begin{center}
\baselineskip= 17 truept

\vspace{.3in}

\noindent
{\Large\bf Black Hole Geometries in Noncommutative String Theory}

\vspace{.4in}

\noindent
{{{\Large Supriya Kar}\footnote{skkar@physics.du.ac.in }}
{{\Large and Sumit Majumdar}\footnote{sumit@physics.du.ac.in}}}

\vspace{.2in}

\noindent
{\large Department of Physics \& Astrophysics\\
University of Delhi, Delhi 110 007, India}

\vspace{.2in}

\thispagestyle{empty}

\baselineskip= 17 truept
\vspace{.6in}
%{\today}

\noindent
{\bf Abstract}: We obtain a generalized Schwarzschild (GS-) and a generalized Reissner-Nordstrom (GRN-)
black hole geometries in ($3+1$)-dimensions, in a noncommutative string theory. In particular, we 
consider an effective theory of gravity on a curved $D_3$-brane in presence of an electromagnetic 
(EM-) field. Two different length scales, inherent in its noncommutative counter-part, are exploited 
to obtain a theory of effective gravity coupled to an $U(1)$ noncommutative gauge theory to all orders 
in $\Theta$. It is shown that the GRN-black hole geometry, in the Planckian regime, reduces to the 
GS-black hole. However in the classical regime it may be seen to govern both
Reissner-Nordstrom and Schwarzschild geometries independently. The emerging notion of $2D$ black holes 
evident in the frame-work are analyzed. It is argued that the $D$-string in the theory may be described 
by the near horizon $2D$ black hole geometry, in the gravity decoupling limit. Finally, our analysis 
explains the nature of the effective force derived from the nonlinear EM-field and accounts for the 
Hawking radiation phenomenon in the formalism.

\baselineskip= 14 truept

\vspace{1in}

%\noindent PACS: 11.25.-w; 11.15.-q

%\noindent Keywords: open string theory, D-branes, noncommutative space-time, Planckian scattering.

%\thispagestyle{empty}%

\end{titlepage}

\baselineskip= 18 truept

%%%%%%%%%%%%%%%%%%%%%%
\section{Introduction}
%%%%%%%%%%%%%%%%%%%%%%

Black holes are known as the classical solutions to the Einstein's general theory of relativity (GTR).
However, the loss of information from a black hole is not governed 
by a pure classical physics. At least, it requires a semi-classical phenomenon, where the space-time 
is treated classically in presence of a quantized matter field in the theory.

\sp 
Very recently, the information loss problem was re-visited by Hawking using a path 
integral formalism \cite{hawking05}. Since such a process involves a decay of pure quantum states 
into mixed ones, it gives rise to a non-unitary quantum theory gravity. In fact, the idea dates back to
the seminal work \cite{hawking1} on Hawking radiation from a black hole. It was established that 
black holes radiate quantum mechanically. The radiation is not exactly thermal in nature, rather 
it arises due to the particle creation at the event horizon of a black hole by an external source for the
matter field \cite{hawking1}-\cite{parikh-wilczek}. 
However there is a caveat in the formalism, as the strong curvature of space-time at Planck
scale is usually ignored to estimate the Hawking radiation. 

\sp
In the context a $D$-brane \cite{polchinski} 
may be seen to provide a natural frame-work to address the problem of
Hawking radiation from a black hole. The idea of strong space-time curvature on the $D_3$-brane world-volume 
is supported by the fact that they are the Planck scale probes in a string theory. 
Interestingly, some of the issues in quantum gravity may be addressed by developing an appropriate 
formalism on a curved $D$-brane. Among the recent developments, the nonlinear electrodynamics on a $D_3$-brane 
\cite{gibbons-hashi}-\cite{gibbons-ishibashi} appears to be a potential candidate to address some of the
semi-classical solutions, such as shock wave and black hole geometries. Above all, several aspects of quantum 
gravity have been addressed on a $D$-brane world-volume in the recent literatures 
\cite{mars-senovilla-vera}-\cite{nasseri}.

\sp
In particular, the noncommutative $D$-brane world-volume \cite{seiberg-witten} may be seen to be 
one such relevant frame-work, to address the Hawking radiation phenomenon and the information loss paradox.
In fact, the closed string decoupling limit in a noncommutative open string theory \cite{minwalla} turns out 
to be a powerful tool as the gravity decouples completely from the theory, but at the Planck scale. 
The cancellation of strong intrinsic curvature in the regime by the strong nonlinearity in the Maxwell 
theory is remarkable. In fact, the extrinsic curvature due to the noncommutative gauge field in the theory 
dominates over that of the gravity. As a consequence
the frame-work allows one to formulate an effective quantum theory of gravity without any ambiguity with the principle 
of equivalence. In particular a $D_3$-brane world-volume incorporating the Einstein's GTR coupled to the nonlinear 
Maxwell's theory may yield a plausible frame-work at Planck scale.
Interestingly, some of the recent field theoretic models \cite{garcia,tamaki}, similar to that of a string, have been 
worked out in the literatures. However the stringy frame-work being natural at Planck scale, provides a wide perspective.

\sp
In this paper, we begin with a $D_3$-brane in presence of an EM-field in a open bosonic string 
theory.{\footnote{The mass-shell condition has been worked out, by one of us in a collaboration \cite{kar-panda},
to show the natural emergence of two different length scales in theory. 
Recently, we have exploited the large and small dimensions in a noncommutative string theory 
\cite{kar-majumdar} to investigate a scattering phenomenon on a $D_3$-brane.
It was shown that the $S$-matrix in the theory generalizes that of a point particle 
scattering in Einstein's theory at Planck scale \cite{thooft}. It was argued that the physical consistency in
quantum gravity imposes a Lorentzian signature on the emerging $D$-string in the theory. 
In fact, the underlying idea of two scales in a noncommutative string theory is precisely in agreement 
with the scaling arguments of Verlinde and Verlinde \cite{ver-2,ver-21}.}} We consider a static gauge condition to
incorporate a nontrivial induced metric on the brane world-volume. 
The brane dynamics is worked out for both ordinary as well as its noncommutative counter-part in the theory. We
obtain a GS- and a GRN-black holes on the $D_3$-brane in the effective theory. 
The mass and the charge of the black holes are shown to receive noncommutative corrections to all order in
$\Theta$. The GRN-black hole geometry is analyzed for various values of its effective parameters 
in the classical and subsequently  in the Planckian regimes. It is shown that
the GRN-geometry reduces to a RN-like geometries in the semi-classical regime. At the other extreme, the GRN geometry
is shown to describe a GS-black hole. In addition, the noncommutative frame-work
gives rise to the notion of a two dimensional black hole in the theory. The Hawking temperature is obtained for the
black holes to analyze the quantum radiation phenomenon. Hawking radiation from the black holes is
explained with the help of the nonlinear ${\bf E}$-field in the theory.

\sp
We plan to organize the paper as follows. In section 2, we describe the set-up to generalize a flat $D_3$-brane to
a curved one. The effective description leading to noncommutative scaling in the frame-work is given in section 3.
In section 4, we perform a series of orthogonal rotations to establish a scaled ``spherical polar''coordinate system.
The generalized black hole geometries are obtained and analyzed in section 5. Exact $2D$ near horizon black holes are 
obtained in section 6. Finally, we conclude with some remarks in section 7.

%%%%%%%%%%%%%%%%%%%%%%%%%%%%%%%%%%%%%%
\section{$D_3$-brane geometry: Set-up}
%%%%%%%%%%$$%%%%%%%%%%%%%%%%%%%%%%%%%%

$Dp$-branes are ($p+1$)-dimensional, short distance, probe in string theory. 
The open bosonic string boundary fluctuations make them dynamical. The closed string background fields, the metric  
${\bar g}_{MN}(X)$ and the simplectic two-form ${\bar{\cal B}}_{MN}(X)$ in the theory, respectively, 
give rise to the induced field $g_{\mu\nu}(Y)= \partial_{\mu}X^M\partial_{\nu}X^N {\bar g}_{MN}$ and 
${\cal B}_{\mu\nu}(Y)=\partial_{\mu}X^M \partial_{\nu}X^N {\bar{\cal B}}_{MN}$ on the brane, 
where $(M,N=1,2,\dots 26)$. The $U(1)$ gauge potential ${\cal A}_M(X)$ at the open string boundary, 
substantially contribute $A_{\mu}(Y)$ on the $D_3$-brane, 
which may be seen due to the $22$-Dirichlet conditions there. Thus all the fields, on the brane world-volume, possess a
nontrivial space-time dependence \cite{kar-1}, which gives rise to the strong curvature in the theory. 
In addition, the $D_3$-brane are charged under the 
Ramond-Ramond forms in type II string theories and may be viewed as nonperturabative extended objects \cite{polchinski}. 
These facts together may imply that a curved world-volume theory provides a viable frame-work to address some of the 
issues in quantum gravity. 

\sp
On the other hand, a construction for the $D$-brane dynamics in a open bosonic string theory has been worked out for 
a constant background two form ${\bar{\cal B}}_{MN}$ field. In addition, the induced metric $g_{\mu\nu}$ has been 
treated as a constant, for simplicity, in the theory. Then, the brane world-volume is described by the 
zero modes of the induced fields. It may appropriately be viewed as the flat $D_3$-brane in the asymptotic limit, 
$i.e.\ r\rightarrow \infty$, of a more general curved brane. In fact, the closed string modes are tangential 
to the $D_3$-brane, which means that string bulk dynamics is not completely decoupled in the theory. 
In addition, in a static gauge condition on the space-time, the nontrivial metric ${\tilde g}_{\mu\nu}=g_{\mu\nu}$. 
Given the facts, it may be possible construct a curved $D_3$-brane dynamics by incorporating the non-zero modes of 
the induced metric in the theory.
Schematically, the space-time curvature as develops on the brane may be represented by a cigar diagram in 
figure 1.
%\ref{sksmf1}
\begin{figure}[ht]
\begin{center}
\vspace*{2.5in}
%\hspace*{-.5in}
\relax{\includegraphics{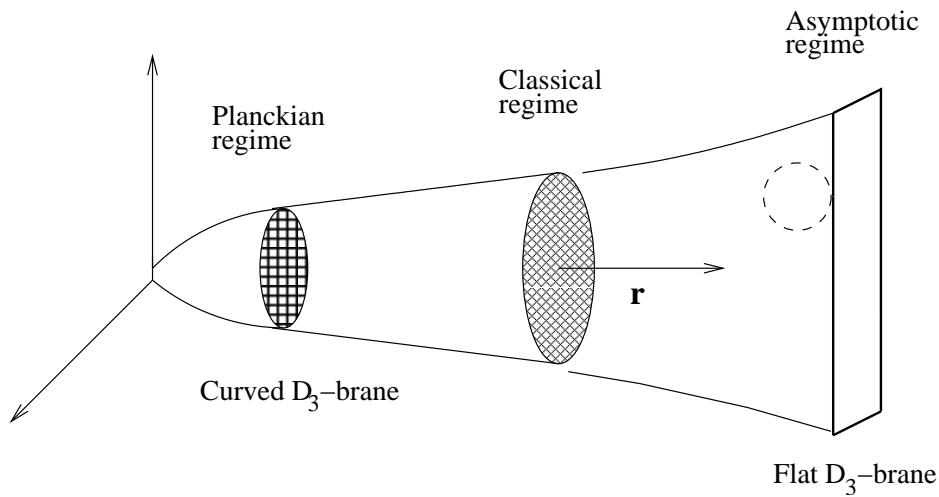}}
\end{center}
\vspace*{-.1in}
\caption{The space-time curvature on the $D_3$-brane in classical and Planckian regimes. The curvature
begins to pierce into the $D_3$-brane world-volume in the 
classical regime itself and gradually develops its way  to the Planckian regime.}
\label{sksmf1}
\end{figure}

%%%%%%%%%%%%%%%%%%%%%%%%%%%%%%%%%%%
\subsection{Asymptotic description}
%%%%%%%%%%%%%%%%%%%%%%%%%%%%%%%%%%%
 
We begin with a flat $D_3$-brane, in presence of a constant two form ${\cal B}$ induced on its world-volume
$(y_1,y_2,y_3,y_4)$. We consider an Euclidean world-volume signature ($+,+,+,+$) all through the paper. 
The Minkowski signature may be obtained by an analytic continuation $y_4\rightarrow it$.
The uniform EM-field on the $D_3$-brane may be expressed by its arbitrary components 
${\bf E}= (0,E_2,E_3)$ and ${\bf B}=(0,B_2,B_3)$. 
Since the induced metric $g_{\mu\nu}$ and the two form ${\cal B}_{\mu\nu}$ are constants on the brane world-volume, 
the DBI-action describes a nonlinear electrodynamics and may be governed by the asymptotic limit of a string theory. 
The brane dynamics is known to be governed by the Dirac-Born-Infeld (DBI) action. It is given by
\bea
S_{\rm DBI}&=&T_D\int d^4y\ \left ({\sqrt{g}} -\ {\sqrt{\big ( g +\bar{\cal B}\big )}}\ \right )\nonumber\\
&=&-{1\over{4g^2_{\rm nl}}}\int d^4y\ {\sqrt{g}}\  g^{\mu\nu}g^{\lambda\rho}\ {\cal B}_{\mu\lambda} {\cal B}_{\nu\rho} 
\ ,\label{act1}
\eea
where $T_D$ denotes the $D_3$-brane tension, $g\equiv\det g_{\mu\nu}$, $(g+ \bar{\cal B})\equiv \det (g +  
{\cal B})_{\mu\nu}$ and $g_{\rm nl}$ is a coupling constant in the nonlinear Maxwell theory.
Though the metric $g_{\mu\nu}=g \delta_{\mu\nu}$ and the two-form ${\cal B}_{\mu\nu}$ are constants, 
its counter-part on the noncommutative $D_3$-brane may precisely be given by a nontrivial effective 
metric \cite{seiberg-witten} 
\be
G_{\mu\nu} = g_{\mu\nu} - \left ( {\cal B}g^{-1}{\cal B}\right )_{\mu\nu}\ .\label{act3}
\ee
Explicitly the effective metric is worked out in, natural units ($\hbar =1=c$), and may be expressed as
\be
ds^2= \left ( g - G^2_N\ g^{-1}{\bf E}^2\right )\left [dy_3^2 + dy_4^2\right ] + 
\left (g+ G^2_N\ g^{-1}{\bf B}^2\right )\left [dy_1^2 + dy_2^2\right ]\ ,\label{act4}
\ee
where $G_N$ denotes the Newton's constant.
The notations used are $E_i=i{\cal B}_{i4}$ for $i=(1,2,3)$ and $B_i= {1\over2} 
\epsilon_{ijk} {\cal B}_{jk}$ is defined with a normalization $\epsilon_{123}=1$. 
In particular, in a typical cartesian coordinate system 
${\cal J}_C$, $i.e.\ (\tau,x,y,z)$, the metric may be re-expressed as
\be
ds^2= \left ( g - G^2_N\ g^{-1}{\bf E}^2\right )\left [d\tau^2 + dz^2\right ] + 
\left (g+ G^2_N\ g^{-1}{\bf B}^2\right )\left [dx^2 + dy^2\right ]\ ,\label{act4c}
\ee
The EM-field may be seen to be governed by an (anti-) parallel configuration.
In the spherical polar coordinate (${\cal J}_S$-) system $(\tau, r, \theta, \phi )$, the 
effective metric becomes  
\be
ds^2= \left ( g - G^2_N\ g^{-1}{\bf E}^2\right )\left [d\tau^2 + dr^2\right ] + 
\left (g+ G^2_N\ g^{-1}{\bf B}^2\right ) r^2 d\Omega^2\ .\label{act41}
\ee
The geometry describes a spherically symmetric spike solution in nonlinear electrodynamics \cite{gibbons-herdeiro}.
The transformation between ${\cal J}_O\rightarrow {\cal J}_S$-coordinates, may also be viewed in two steps. First
the ${\cal J}_S$-coordinates are obtained from that in ${\cal J}_O$-system by the map
\bea
&&\tau\rightarrow \tau\; ,\qquad z= r \cos\theta\ ,\nonumber\\
&&x= r \sin\theta \cos\phi\;\ {\rm and}\quad y=r \cos\theta \sin\phi\ .\label{act42}
\eea 
In the second step, an orthogonal rotation ${\cal R}_{\theta}$ of ($r,\theta$)-plane in ${\cal J}_S$-system, 
by an arbitrary angle $\theta$, around the symmetry axes ($\tau,\phi$) leads to the effective metric (\ref{act4}).

\sp
Performing two successive orthogonal rotations in anti-clock-wise directions, respectively, around $\hat\phi$ axis 
by an angle $\pi/2$ and around $\hat r$ by $\pi$ in ${\cal J}_S$-system, one finds  
\be
\hat\tau \rightarrow \hat\tau\ ,\qquad 
\hat r \rightarrow \hat\theta\ ,\qquad \hat\theta \rightarrow \hat r \qquad {\rm and}\quad \hat\phi 
\rightarrow - \hat \phi\ .\label{act43}
\ee
In the new coordinate system, the effective metric (\ref{act4}) becomes
\bea
ds^2&=&\left ( g- G^2_N\ g^{-1}{\bf E}^2\right )d\tau^2 + \left (g+ G^2_N\ g^{-1}{\bf E}^2\right ) dr^2 
\nonumber\\
&&\qquad\qquad\qquad\quad\ + \left ( g- G^2_N\ g^{-1}{\bf B}^2\right ) r^2 d\theta^2 + 
\left (g+ G^2_N\ g^{-1}{\bf B}^2\right ) r^2 \sin^2\theta\ d\phi^2 \ .\label{act44}
\eea
The orthogonal transformations, between different coordinate systems, show that
different geometries (\ref{act4c}), (\ref{act41}) and (\ref{act44}) may directly be obtained from 
the eq.(\ref{act4}) by an appropriate choice of orthogonal axes in the theory. For instance, 
the $EM$-field breaks the spherical symmetry $S_2\rightarrow S_2^{\Theta}$ in the new coordinates (\ref{act44}), 
which is otherwise preserved in (\ref{act41}). In other words, the noncommutative geometry seems to be 
generated in the new coordinates. We shall see that the effective metric components 
(\ref{act44}) receive all order $\Theta$ corrections in the theory.

%%%%%%%%%%%%%%%%%%%%%%%%%%%%%
\subsection{Classical regime}
%%%%%%%%%%%%%%%%%%%%%%%%%%%%%

The asymptotic description (\ref{act1}) may be generalized to include a slow variation 
in the metric $g_{\mu\nu}$ in the classical regime. Then the $D_3$-brane dynamics 
would be governed by coupling Einstein's theory to an appropriate DBI-action. 
However, we consider a static gauge condition on space-time to begin with. Then the bulk metric may be viewed 
on the world-volume and the complete action becomes 
\be
S= {1\over{16\pi G_N}}\int d^4y \ {\sqrt{g}}\ R \ +\ S_{\rm DBI}\ ,
\label{act5}
\ee
where $R$ is the scalar curvature in the theory. The expansion under the square-root in the action 
(\ref{act1}), in the regime, is worked out. Then the action (\ref{act5}) becomes
\be
S= \int d^4y\ {\sqrt{g}}\ \left (\ {1\over{16\pi G_N}}\ R \ - 
{1\over4} g^{\mu\nu}g^{\lambda\rho}\ {\cal F}_{\mu\lambda} {\cal F}_{\nu\rho} \ + \ 
{\cal O}({\cal F}^4)\ +\ \dots\ \right )\ .\label{act6}
\ee
The $U(1)$ gauge invariant field strength in the theory is governed by ${\bar{\cal F}}_{\mu\nu}=
({\cal B} + 2\pi\alpha'{\bar F})_{\mu\nu}$. 
The higher order terms in gauge field may be ignored in the classical regime. Then, the nonlinear action (\ref{act6}) 
resembles to the Einstein's theory coupled to the Maxwell's. The equation of motion for the gauge field becomes
\be
\partial_{\mu}{\cal F}^{\mu\nu}=0\ .\label{act2}
\ee
In addition, the metric variation in (\ref{act6}) leads to 
\be
R_{\mu\nu} -{1\over2}g_{\mu\nu}R = \left (8\pi G_N\right ) T_{\mu\nu}\ ,\label{act7}
\ee 
where the energy momentum tensor is given by
\bea
T_{\mu\nu}&=&{1\over{\sqrt{g}}} {{\delta S_{\rm DBI}}\over{\delta g^{\mu\nu}}}\ ,\nonumber\\ 
&=&{1\over2}\left ( {1\over4} g_{\mu\nu}{\cal F}_{\mu'\nu'}{\cal F}^{\mu'\nu'}
- {\cal F}_{\mu\lambda}{\cal F}^{\lambda}_{\nu}\ \right )\ .\label{act8}
\eea
The gauge potential $A_{\mu}=(A_{\tau}$, $A_r$, $A_{\theta}$, $A_{\phi})$ is governed by the 
equations of motion (\ref{act2}) and is given by
\be
A_{\mu} = \left ( {{-iQ_e}\over{r}},\ 0,\ 0,\ Q_m\cos \theta\right )\ ,\label{act81}
\ee
where $Q_e$ and $Q_m$ are constants, respectively, denote the electric and magnetic charges in the theory.
Since the components of the gauge potential (\ref{act81}) remain unchanged in their magnitude under the rotations 
(\ref{act43}), the EM-field is worked out in the transformed coordinate frame.
Then the EM-field may be seen to be governed by an anti-parallel 
configuration
\be
{\bf E}= -{{Q_e}\over{r^2}}\ {\hat r}\qquad\quad {\rm and} \qquad {\bf B}= {{Q_m}\over{r^2}}\ {\hat r}\ .\label{act82}
\ee
To leading order in nonlinearity, $i.e.$ for ${\cal B}=0$, 
the theory (\ref{act6}) precisely reduces to the Einstein's theory coupled to the Maxwell's.
In presence of both the charges, the geometry may be seen to be
governed by 
\bea
ds^2&=&\left (1-{{2G_NM}\over{r}} + {{G_N Q_e^2}\over{r^2}} \right ) d{\tau}^2 
+\left (1- {{2G_NM}\over{r}}+ {{G_N Q_e^2}\over{r^2}} \right )^{-1} dr^2 
\nonumber\\
&&\qquad\qquad\qquad\quad +\left (1 -\ {{G_N Q_m^2}\over{r^2}} \right ) 
r^2 d\theta^2\ +\ \left (1 -\ {{G_N Q_m^2}\over{r^2}} \right )^{-1} r^2\sin^2\theta\ d\phi^2\ .\label{act9}
\eea
For $Q_m=0$, the solution restores the spherical symmetry and governs the Reissner-Nordstrom (RN-) black hole geometry 
in the regime. Interestingly, some  generalizations of RN-geometry, in the classical regime, have been studied 
\cite{garcia,tamaki} in a different context. 

\sp
On the other hand, in the classical regime $F_{\mu\nu}\neq 0$. Then, the expression for the effective metric (\ref{act3}) becomes
\be
G_{\mu\nu}= g_{\mu\nu} - \left ({\bar{\cal F}}g^{-1}{\bar{\cal F}}\right )_{\mu\nu} \ +\ {\cal O}({\cal F}^4)\ +\ \dots\ \ .
\label{act33}
\ee
Since $T_{\mu\nu}$ is weak, one may approximate the gravitational solution in the theory
by the Schwarzschild black hole geometry. A semi-classical solution to the equations of motion 
(\ref{act2}) and (\ref{act7}) is worked out in the classical regime. 
Ignoring the higher order terms in the modified metric (\ref{act33}), we obtain 
\bea
ds^2&=&\left (1-{{2G_N M}\over{r}}\right ) \left ( 1 - {{G^2_NQ_e^2}\over{r^4}} \right ) d{\tau}^2 
+\left (1- {{2G_N M}\over{r}}\right )^{-1}\left (1 - {{G^2_N Q_e^2}\over{r^4}} \right )^{-1} dr^2 
\nonumber\\
&&\qquad\quad +\left (1 -\ {{G^2_N Q_m^2}\over{r^4}} \right ) 
r^2 d\theta^2\ +\ \left (1 -\ {{G^2_N Q_m^2}\over{r^4}} \right )^{-1} r^2\sin^2\theta\ d\phi^2\ .\label{act10}
\eea
The generalized classical solution retains spherical symmetry when $Q_m=0$. It possesses a 
curvature singularity at $r=0$. On the other hand, the coordinate singularity turns out to become a scale
dependent one and occurs either at $r_c={\sqrt{G_NQ_e}}$ or at $r_s=2G_NM$, which leads to a Schwarzschild  
black hole in the theory.  The fact that the theory (\ref{act6}) allows two 
semi-classical black holes (\ref{act9}) and (\ref{act10}) for ${\cal B}=0$ may lead to an apparent puzzle. 
We shall see that the puzzle may be resolved by taking into account the 
${\cal B}$-field in the formalism.  The nonlinearity in the $U(1)$ gauge sector will be seen to modify 
its charges
\be
Q_{\rm eff}^2 = Q^2_e \left ( 1 - {{G_N |{\bf E}|}\over{Q_e}}\ - \ \dots\ \right )\qquad {\rm and}\qquad
{\bar Q}_{\rm eff}^2 = Q^2_m \left ( 1 - {{G_N} |{\bf B}|\over{Q_m}}\ - \ \dots\ \right )
\ .\label{act11}
\ee 
Then, these two different semi-classical black hole geometries may be obtained from a generalized RN-black hole
in a noncommutative string theory. For instance, the leading order term in eq.(\ref{act11}) would 
appropriately represent the RN-like black hole geometry (\ref{act9}) and the 1st order term there would govern the 
Schwarzschild like geometry (\ref{act10}). We postpone the detail of computations on the corrections to a later 
section 5.

%%%%%%%%%%%%%%%%%%%%%%%%%%%%%
\subsection{Planckian regime}
%%%%%%%%%%%%%%%%%%%%%%%%%%%%%

In the case, the space-time curvature on the $D_3$-brane becomes significant due to the nontrivial induced metric and 
gauge fields.  The appropriate world-volume dynamics on a curved $D_3$-brane may be 
described at the Planck scale by an appropriate generalization of the action (\ref{act5}).
The expansion under the square-root in the action (\ref{act1}) is performed by keeping track of a 
particular order in gauge field strength (say ${\cal F}^4$). Then, the complete action 
may be simplified to yield
\bea
S&=& \int d^4y\ 
{\sqrt{g}}\ \left (\ {1\over{16\pi G_N}}\ R \ - 
{1\over4} \Big [\ {\cal F}^2 
-\ {1\over2} {\cal F}{\cal F}_{+}\ F_{-}^2\ K^2\big ( {\cal F}\big )\ \Big ]\ \right )\nonumber\\
&=& \int d^4y\ {\sqrt{g}}\ \Bigg (\ {1\over{16\pi G_N}}\ R \ - {1\over4} \Big [\ {\cal F}^2 
-\ {1\over2}{\cal F}^4\ K^2\big ({\cal F}\big )\nonumber\\
&&\;\ \qquad\qquad\qquad\qquad\qquad\qquad\qquad + {1\over2}\Big ( {\cal F}^3\ {}^{\star}{\cal F} 
- {\cal F}\ {}^{\star}{\cal F}^3 + \big [{\cal F}\ {}^{\star}{\cal F}\big ]^2\Big )
K^2\big ({\cal F}\big )\Big ]\Bigg )\ ,\label{action113}
\eea
where $G_N= l^2_p$ signify the Planck scale in the theory.
The field strengths
${\cal F}_{\pm}= ({\cal F}\pm {}^{\star}{\cal F})$ and $K\big ({\cal F}\big )$ contains all higher order terms in 
field strength.
The Hodge dual of ${\cal F}$ is denoted as ${}^{\star}{\cal F}$. Explicitly,
\be
{}^{\star}{\cal F}^{\mu\nu} = {1\over{2\sqrt{g}}}{\cal E}^{\mu\nu\rho\lambda} \ {\cal F}_{\rho\lambda}
\ ,\label{action114}
\ee
where ${\cal E}^{\mu\nu\rho\lambda}$ is a covariant antisymmetric tensor density.
However, the Minkowski's inequality  in the theory (\ref{action113}) gives rise to the (anti-) self-duality condition 
\be
{\cal F}_{\mu\nu} = \pm^{\star}{\cal F}_{\mu\nu}\ .\label{action115}
\ee
It implies $E_i=B_i$ in the theory. In addition, the self duality can be seen to impose a lower bound on the DBI-action
\cite{gibbons-hashi}.
Since all the higher order terms in gauge fields vanish, the theory (\ref{action113}) yields an 
exact stringy description. Then the relevant action, on a curved
$D_3$-brane, becomes
\be
S= \int d^4y\ {\sqrt{g}}\ \Big (\ {1\over{16\pi G_N}}\ R  \ - 
{1\over4} g^{\mu\lambda} g^{\nu\rho}\ {\cal F}_{\mu\nu}{\cal F}_{\lambda\rho}\ \Big )\ .\label{action116}
\ee
The Einstein's equations of motion (\ref{act7}), in the regime may be seen to be governed by the vacuum equations 
$i.e.\ T_{\mu\nu} = 0$. This is  due to the self-duality condition on the nonlinear gauge field (\ref{action115}) 
in the theory. In other words, the Einstein's equation is not modified by the presence of a nonlinear Maxwell 
term on a curved $D_3$-brane. The remaining equation of motion is that for the gauge field and is given by
\be
D_{\mu}{\cal F}^{\mu\nu}=0\ ,\label{action118}
\ee
where $D_{\mu}\equiv \big (\partial_{\mu}-{1\over2}\partial_{\mu}g^{\lambda\rho} g_{\lambda\rho}\big )$.
For convenience, we postpone the geometric detail on the $D_3$-brane to an appropriate section 5.

%%%%%%%%%%%%%%%%%%%%%%%%%%%%%%%%%%%%%%%%
\section{Effective noncommutative frame}
%%%%%%%%%%%%%%%%%%%%%%%%%%%%%%%%%%%%%%%%

%%%%%%%%%%%%%%%%%%%%%%%%%%%%%%%%%%%%%%%%%%%%%%%%%%%%%%%%%%%%%%%%%%%
\subsection{Exact action in a $4$-dimensional Euclidean space-time}
%%%%%%%%%%%%%%%%%%%%%%%%%%%%%%%%%%%%%%%%%%%%%%%%%%%%%%%%%%%%%%%%%%%

In this section, we focus on the relevant noncommutative dynamics corresponding to a curved $D_3$-brane action 
(\ref{action116}). However, we begin with a flat $D_3$-brane as described in section 2.1.
It is known that Seiberg-Witten map \cite{seiberg-witten} transforms the ordinary gauge theory on the 
$D_3$-brane (\ref{act1}) to a noncommutative one. Though the induced fields 
$g$ and ${\cal B}$ are constants, the invariant field strength ${\cal F}$ modifies the Einstein's metric 
non-trivially{\footnote{Strictly, speaking the electromagnetic field modifies the string metric, which is related 
to the Einstein's metric by a conformal factor.}}. 
The flat brane picks up noncommutative geometry and is governed by an effective metric $G_{\mu\nu}$ (\ref{act33}). 
Then, the noncommutative $D_3$-brane dynamics may be obtained from the DBI-action (\ref{act1}) 
with appropriate modifications. With Riemannian signature ($\mu ,\nu= 1,2,3,4$), the relevant effective action becomes 
\be
{\hat S}_{\rm DBI}= T_D^{\rm nc}\int d^4y\ \left ( {\sqrt{G}} - \ {\sqrt{G + 2\pi\a' {\hat F}}}\ \right )\ ,\label{action2}
\ee
where $G\equiv\det G_{\mu\nu}$. The noncommutative $U(1)$ field strength on the world-volume is
given by 
\bea
{\hat F}_{\mu\nu}&=&\pr_{\mu}{\hat A}_{\nu} - \pr_{\nu}{\hat A}_{\mu} - i\big 
[{\hat A}_{\mu}\ ,\ {\hat A}_{\nu}\big ]_{\star}\nonumber\\
&=&\pr_{\mu}{\hat A}_{\nu} - \pr_{\nu}{\hat A}_{\mu}\ + \Theta^{\rho\lambda}\ 
\partial_{\rho}{\hat A}_{\mu}(x) \partial_{\lambda} {\hat A}_{\nu}(y){\big |}_{x=y} +\ {\cal O}(\Theta^2) \ ,
\label{action3}
\eea
where $\Theta$ is a noncommutative parameter in the theory. The effective theory on the noncommutative $D_3$-brane may be 
described similar to its counter-part (\ref{act5}) with ordinary geometry. 
The relevant action for a curved $D_3$-brane becomes
\bea
{\hat S}&=&{1\over{16\pi G_N}}\int d^4y \ {\sqrt{G}}\ {\cal R} \ +\ {\hat S}_{\rm DBI}
\nonumber\\
&=& \int d^4y\ {\sqrt{G}}
\left ( {1\over{16\pi G_N}} {\cal R}\ -  {1\over4} 
Tr \Big ({\hat F}_{\mu\nu} {\hat F}^{\mu\nu}\Big )\ -
{\cal O}({\hat F}^4)\ +\ \dots \ \ \right )\ .\label{action4}
\eea
The higher order terms starting with ${\cal O}({\hat F}^4)$ essentially account for the stringy corrections in the theory. 
They may be worked out explicitly, using the Seiberg-Witten map, to yield a relation between the 
noncommutative $U(1)$ gauge field and its ordinary counter-part. It is given by 
\be
{\hat A}_{\mu} = A_{\mu} -{1\over2} \Theta^{\nu\lambda} A_{\nu}\left (\partial_{\lambda}A_{\mu} + 
{\cal F}_{\lambda\mu}\right ) +\ {\cal O}(\Theta^2)\ .\label{black211}
\ee
The corresponding field strengths are related as  
\be
{\hat F}_{\mu\nu} = {\cal F}_{\mu\nu} + \Theta^{\rho\lambda} \left ( {\cal F}_{\mu\rho}{\cal F}_{\nu\lambda} - 
A_{\rho}\partial_{\lambda}{\cal F}_{\mu\nu}\ \right )\ + \ {\cal O}(\Theta^2)\ .\label{black212}
\ee
Then, the gauge field corrections to the world-volume dynamics (\ref{action4}) may be computed.  
To ${\cal O}(\Theta)$, it may be expressed as
\bea
{\hat F}^4 + {\rm (higher\ orders)} &\equiv&- {{K^2({\cal F})}\over2} \Bigg 
[ {\cal F}_{\mu\nu}{\cal F}_+^{\mu\nu}\ {\cal F}_-^2 
+ \Theta^{\lambda\rho} \Big ({\cal F}_{\mu\lambda}{\cal F}_{\nu\rho} {\cal F}_+^{\mu\nu}\ {\cal F}_-^2\qquad
\qquad\qquad\qquad\nonumber\\ 
&&\qquad\qquad\quad - A_{\lambda}\partial_{\rho}{\cal F}_{\mu\nu}\ {\cal F}_+^{\mu\nu}\ {\cal F}_-^2\ +\  
({\cal F}_+)_{\mu\lambda}({\cal F}_+)_{\nu\rho}\ {\cal F}^{\mu\nu} \ {\cal F}_-^2\nonumber\\
&&\qquad\qquad\qquad\; + 2({\cal F}_-)_{\mu\lambda} ({\cal F}_-)_{\nu\rho}\ ({\cal F}_-)^{\mu\nu}\ 
{\cal F}_{\delta\sigma}({\cal F}_+)^{\delta\sigma} + \dots\ \Big ) \Bigg ] .\label{black2121}
\eea
The higher order terms show the presence of strong gauge curvature in the theory (\ref{action4}). 
Using self duality condition for the ordinary gauge field ${\cal F}_-=0$, it is straight-forward to see 
that all the higher order terms vanish identically in a particular combination of the gauge field. 
Then the effective action (\ref{action4}) becomes exact to all orders in stringy corrections. It is given by
\be
{\hat S}=\int d^4y\ {\sqrt{G}} \left ( {1\over{16\pi G_N}} {\cal R}\ -  {1\over4} 
G^{\mu\lambda} G^{\nu\rho}\ {\hat F}_{\mu\nu} \star {\hat F}_{\lambda\rho}\ \right )\ .\label{action42}
\ee
The Moyal $\star$-product in the noncommutative gauge theory is known to introduce nonlocal terms in the action due to the
infinite number of derivatives there. 
The equations of motion, respectively, for the effective metric $G_{\mu\nu}$ and the noncommutative gauge potential
${\hat A}_{\mu}$ are given by
\bea
&&{\cal R}_{\mu\nu} -{1\over2}G_{\mu\nu}{\cal R} = \left (8\pi G_N\right ) {\hat T}_{\mu\nu}
\nonumber\\
{\rm and}\qquad &&\ \left [\delta^{\lambda}_{\mu} - \Theta^{\lambda\rho}\ 
\partial^y_{\rho}{\hat A}_{\mu}(y)\ +\ {\cal O}(\Theta^2)\ \dots\  
\right ]_{y\rightarrow x}{\cal D}_{\lambda}{\hat F}^{\mu\nu}(x)=0
\ ,\label{action421}
\eea 
where ${\hat T}_{\mu\nu}$ is the energy momentum tensor and 
${\cal D}_{\lambda}= \big ( \partial_{\lambda}-{1\over2} \partial_{\lambda}G^{\rho\delta} G_{\rho\delta}\big )$. 
We shall see that ${\hat T}_{\mu\nu}\neq 0$ in the effective theory (\ref{action42}), which is unlike to its 
ordinary counter-part (\ref{action116}). 

\sp
On the other hand, the vanishing stringy contributions (\ref{black2121}) may imply that the noncommutative theory 
(\ref{action42}) naturally corresponds to the gravity decoupling regime at Planck scale.{\footnote{It is known that the
${\bf E}$-string drains off the stringy contribution and leads to a tension-less string in the effective theory.}} 
Then the ${\bf E}$-field in its ordinary
counter-part is equivalently described by ${\bf E}\rightarrow {\bf E}_c$. Since all the closed string modes decouple 
from the theory, the effective metric becomes 
$G_{\mu\nu} \rightarrow - (2\pi\alpha')^2[{\cal F}_{\mu\lambda}\delta^{\lambda\rho}{\cal F}_{\rho\nu}]$. 
Thus the effective dynamics, in the decoupling regime, is completely governed by the nonlinear 
electromagnetic field in the theory. 

\subsection{Relevant modes of quantum gravity}

In the effective theory of gravity on the $D_3$-brane, the noncommutative parameters ($\Theta_E$ and $\Theta_B$) are
fixed at any given energy scale in the theory. They take nonzero values in a wide range of energy including
the classical regime. Since the radius of time-like brane coordinate is of the order of string length $l_s$ 
\cite{kar-panda}, the noncommutativities $[y^0,y^{2,3}]=i\Theta_E^{2,3}$ imply large length scales, of the order of 
$l_{\perp}=|\Theta_E|/l_s$, along $y^{2,3}$. Similarly $[y^2,y^1]= -i\Theta_B^3$ and $[y^3,y^1]=i\Theta_B^2$ would 
enforce a string scale along $y^1$-coordinate in the theory.
Two different length scales $l_s$ and $l_{\perp}$ in the noncommutative frame-work \cite{kar-majumdar}, allow one to scale
the $4$-dimensional effective theory in terms of a parameter $\lambda= {l_s}/{l_{\perp}}<< 1$. In other words, 
the noncommutative scaling is appropriately described by the limit $\lambda\rightarrow 0$. 
Since the induced fields ($g,{\cal B}$) are dimensionless in the frame-work, the scaling naturally introduces 
large and small dimensions in the theory. Under the scaling, the transverse ($\perp$-) coordinates $y^i\rightarrow y^i$ 
and the longitudinal ($L$-) ones $y^{\alpha}\rightarrow \lambda y^{\alpha}$ for ($\alpha = 1,4$ and $i=2,3$).
Thus within the geometrical setup, it is natural to consider a gauge choice $G_{i\alpha}=0$ for the effective metric, 
$i.e.$
\be
G_{\mu\nu}= \pmatrix { {{\bar h}_{\alpha\beta}} & {0} \cr {0} & {h_{ij}} }\ . \label{action5} 
\ee
${\bar h}_{\alpha\beta}$ and $h_{ij}$ represent the metric components, respectively, on the $L$- and $\perp$-spaces
It is important to note that though the zero modes on the $D_3$-brane world-volume are noncommutative, 
its $L$- and $\perp$-spaces are independently described by the ordinary geometry, 
$i.e.\ \Theta^{\alpha\beta}=0= \Theta^{ij}$.
 
\sp
\noindent
In the gauge (\ref{action5}), the effective action (\ref{action4}) may be re-expressed in terms of 
a noncommutative scaling parameter $\lambda$ \cite{ver-2,ver-21,kar-maharana}. It is given by
\bea
{\hat S}_{\rm eff}&=&\int d^2y^{(\alpha)}\ d^2y^{(i)} 
{\sqrt{\bar h}}\ {\sqrt{h}}\ \Big [\ {1\over{16\pi}}{\cal R}_h + {1\over{64\pi}}\ h^{ij}\ \partial_i
{\bar h}_{\alpha\beta} \partial_j{\bar h}_{\gamma\delta} \epsilon^{\alpha\gamma}\epsilon^{\beta\delta}\nonumber\\
&&\qquad\qquad\qquad\qquad\qquad\;\;\;  +\ {1\over{16\pi\lambda^2}}\Big ( {\cal R}_{\bar h} + 
{1\over4}\ {\bar h}^{\alpha\beta}\ 
\partial_{\alpha}{h}_{ij} \partial_{\beta}{h}_{kl} \epsilon^{ik}\epsilon^{jl}\Big )\nonumber\\
&&\qquad\qquad\qquad\qquad\qquad\;\;\;  -\ {1\over4} 
\Big ( {1\over{\lambda^2}}\ {\bar h}^{\alpha\beta} {\bar h}^{\gamma\delta}\ 
{\hat F}_{\alpha\gamma} {\hat F}_{\beta\delta}\ + {\lambda^2}\ {\bar h}^{ij} h^{kl}\ {\hat F}_{ik} 
{\hat F}_{jl} \nonumber\\
&&\qquad\qquad\qquad\qquad\qquad\qquad\qquad\qquad\qquad\qquad\qquad +\  
2 {\bar h}^{\alpha\beta} h^{ij}\ {\hat F}_{\alpha i}\star 
{\hat F}_{\beta j} \Big )\ \Big ] \ .\label{action6}
\eea
In the scaling limit, $\lambda\rightarrow 0$, the on shell action becomes
\be
{\hat S}_{\rm eff}=\int d^2y^{(\alpha)} d^2y^{(i)} {\sqrt{\bar h}}\ {\sqrt{h}}\; 
\Big [ {1\over{16\pi}}{\cal R}_h + {1\over{64\pi}}\ h^{ij}\ \partial_i
{\bar h}_{\alpha\beta} \partial_j{\bar h}_{\gamma\delta} \epsilon^{\alpha\gamma}\epsilon^{\beta\delta}
- {1\over2} {\bar h}^{\alpha\beta} h^{ij}\ {\hat F}_{\alpha i}\star 
{\hat F}_{\beta j}\ \Big ) \Big ] \ .\label{action7}
\ee
The action is obtained by using the vacuum field configurations
\bea
&&\partial_{\alpha}h_{ij}= 0\ ,\nonumber\\
&& {\cal R}_{\bar h} = 0\nonumber\\
{\rm and}\quad &&{\hat F}_{\alpha\beta}=0\ .\label{action8}
\eea
As pointed out, since the noncommutativity in the frame-work is utilized to obtain the small and large length 
scales, both $L$- and $\perp$-spaces in the frame-work are described by ordinary geometries. 
The general solutions to the equations of motion are given by
\bea
&&h_{ij}\equiv h_{ij}\big ( y_{\perp}\big )\ ,\nonumber\\
&&{\tilde h}_{\alpha\beta}\equiv \partial_{\alpha}X^a\partial_{\beta}X^b\ \delta_{ab}\nonumber\\
{\rm and}\quad &&{\hat A}_{\alpha}=0\ ,\label{action9}
\eea
where $X^a(y_{\perp})$ are arbitrary brane modes on a projected space for 
$a=( 1,2,3,4)$. In a static gauge ${\bar h}_{\alpha\beta} \rightarrow \delta_{\alpha\beta}$.

\sp
We learned that the $4$-dimensional effective action (\ref{action7}) governs the 
gravitational dynamics in the $\perp$-space 
and that for the matter field in the $L$-space. The noncommutative matrix parameter $\Theta^{\mu\nu}$ 
in the frame-work act as a propagator in the internal space-time connecting the $L$- and $\perp$-spaces.
The relevant equations of motion described by the $\perp$-space are given by
\bea
&&({\cal R}_h)_{ij} -{1\over2}h_{ij} {\cal R}_h - {1\over{16}}\left ( h_{ij} h^{kl}\partial_k{\bar h}_{\alpha\beta}
\ \partial_l{\bar h}_{\gamma\delta} - 2\partial_i h_{\alpha\beta}\ \partial_j{\bar h}_{\gamma\delta}\right ) 
\epsilon^{\alpha\gamma}\epsilon^{\beta\delta} = (8\pi) {\hat T}_{ij}\,
\qquad\nonumber\\
{\rm and}&&\left [ \delta^{\lambda}_{\alpha} - \Theta^{\lambda k}\partial^y_k 
{\hat A}_{\alpha}(y)\right ]_{y\rightarrow x} {\cal D}^{\perp}_{\lambda} {\hat F}^{\alpha i}(x) = 0
\ .\label{action91}
\eea
Similarly for the $L$-space, the equations of motion are given by
\bea
&&{1\over2} {\bar h}_{\alpha\beta} \partial_i{\bar h}_{\lambda\rho}\ \partial^i{\bar h}_{\gamma\delta}
\ \epsilon^{\lambda\gamma}\epsilon^{\rho\delta} +  \left (
h_{kl}\ \partial_i h^{kl}\ \partial^i {\bar h}_{\gamma\delta} + 2 \partial_i\partial^i {\bar h}_{\gamma\delta} \right )
\epsilon^{\gamma}_{\alpha}\epsilon^{\delta}_{\beta} = (64\pi) {\hat T}_{\alpha\beta}
\qquad\nonumber\\
{\rm and}&&\left [ \delta^k_i - \Theta^{k\lambda}\partial^y_{\lambda} 
{\hat A}_i(y)\right ]_{y\rightarrow x} {\cal D}^{\parallel}_k {\hat F}^{\alpha i}(x) = 0
\ .\label{action92}
\eea
Using the general nature of the solutions (\ref{action9}), the equations of motion for 
${\hat A}_i$ simplifies drastically to yield
\be
\ {\cal D}^{\perp}_{\alpha}\partial^{\alpha} {\hat A}^i = 0\ .\label{action10}
\ee
The energy momentum tensor (\ref{act8}) is computed in the noncommutative frame-work to yield $T_{\alpha i}=0$. 
The remaining non-vanishing components turn out to yield
\bea
&&{\hat T}_{\alpha\beta} = {1\over2}\ \Bigg [ {1\over2}{\bar h}_{\alpha\beta}\ h^{ij}\ 
{\hat F}_{\alpha'i}\star{\hat F}^{\alpha'}_j\ -\ h^{ij}\ {\hat F}_{\alpha i}\star{\hat F}_{\beta j}\ \Bigg ]
\nonumber\\
{\rm and }\quad &&{\hat T}_{ij}= {1\over2}\Bigg [ {1\over2} h_{ij}\ h^{kl}\ 
{\hat F}_{\alpha k}\star{\hat F}^{\alpha}_l\ -\ {\bar h}^{\alpha\beta}\ {\hat F}_{\alpha i}\star{\hat F}_{\beta j}\ \Bigg ]
\ .\label{action11}
\eea
Since the $\perp$-components ${\hat T}_{ij}$ are weak in the frame-work, the energy momentum tensor in the theory
may well be approximated by ${\hat T}_{\alpha\beta}$. It is straight-forward to
check that ${\hat T}_{\alpha\beta}$ is traceless. Being a symmetric tensor, ${\hat T}_{\alpha\beta}$ is survived by 
two degrees of freedom in the theory. They may be identified with the left and right moving momenta of the 
noncommutative strings \cite{kar-majumdar}.  Using eq.(\ref{black212}), the relevant energy momentum tensor 
may be expressed in terms of the ordinary gauge field strength. The self-duality condition (\ref{action115}) 
among the gauge fields in the theory simplifies the expression for ${\hat T}_{\alpha\beta}$. 
To ${\cal O}(\Theta)$, it becomes
\bea 
&&{\hat T}_{\alpha\beta} = {1\over2}\ \Theta^{k\lambda}\Big [\ {\cal F}_{\lambda i}
\Big ({\bar h}_{\alpha\beta}\ {\cal F}^{i\rho}{\cal F}_{\rho k}\ + \ 2{\cal F}^i_{\alpha}{\cal F}_{k\beta}\ \Big ) 
\qquad\qquad\qquad \nonumber\\
&&\qquad\qquad\qquad\qquad\qquad\qquad\qquad +\
A_{k} \Big ({\bar h}_{\alpha\beta}\ {\cal F}_{\rho i}\ \partial_{\lambda}{\cal F}^{i\rho}\ -\ 
2{\cal F}_{\alpha i}\ \partial_{\lambda}{\cal F}^i_{\beta}\ \Big )\ \Big ]\ .\label{action12}
\eea
It confirms that the ${\hat T}_{\mu\nu}$ contributes significantly in the noncommutative frame-work.
Since the $L$- and $\perp$-spaces are described independently by ordinary geometries in the effective
theory, $T_{\mu\nu}$ and $T_{ij}$ are traceless and hence $T_{{\tilde\tau}{\tilde\tau}}=-T_{{\tilde r}{\tilde r}}$.

%%%%%%%%%%%%%%%%%%%%%%%%%%%%%%%%%%%%%%%%%%%%%%%%%%%%%%%%%%%%%%%%%%%%%%%%%%%%%%%%%%%%%%%%%%%%%%%%%%%%%%%%%%%%%%%%
\section{Noncommutative scaling in ($\tilde\tau,{\tilde r},{\tilde l}_{\theta},{\tilde l}_{\phi}$)-coordinates}
%%%%%%%%%%%%%%%%%%%%%%%%%%%%%%%%%%%%%%%%%%%%%%%%%%%%%%%%%%%%%%%%%%%%%%%%%%%%%%%%%%%%%%%%%%%%%%%%%%%%%%%%%%%%%%%%

The scaling analysis discussed in the previous section is a consequence of space-time noncommutativity in the 
theory. Strictly speaking, it makes sense in the cartesian coordinate system. We use tilde notation in a noncommutative
frame-work to distinguish them from their (ordinary) counter-part in the usual coordinates. For instance, 
we denote the noncommutative cartesian coordinates,
$(\tilde\tau,\tilde x,\tilde y,\tilde z)$ with Riemannian signature on the $D_3$-brane by ${\tilde{\cal J}}_C$-system.
The limit $\Theta\rightarrow 0$, implies ${\tilde{\cal J}}_C\rightarrow {\cal J}_C$.
On the other hand, it may be possible to approximate the ``spherical polar'' coordinates 
$(\tilde\tau,\tilde r,\tilde\theta, \tilde\phi)$ in a noncommutative set-up, $i.e.\ {\tilde{\cal J}}_S$-system.
In particular, the noncommutative constraints derived from the ${\tilde{\cal J}}_C$-system may appropriately be
incorporated in the ${\tilde{\cal J}}_S$-system.

\sp
\noindent
Since the $L$- and $\perp$-spaces are described by two independent length scales, 
the $4$-dimensional space-time may be approximated by two independent ($2\times 2$)-blocks, say 
($\tilde\tau,\tilde z$) and ($\tilde x,\tilde y$). In particular, from a $4$-dimensional point of view,
the ($\tilde\tau,\tilde z$) space may be viewed with a fixed polar angle $\tilde\theta$ and then 
the remaining ($\tilde x,\tilde y$) space may be described with a fixed radial coordinate $\tilde r$.
A priori, two sub-spaces in the theory are consistent with the total number space-time degrees of freedom.
However, a careful analysis reveals that the noncommutative constraints further freeze some degrees of freedom
in the theory.

%%%%%%%%%%%%%%%%%%%%%%%%%%%%%%%%%%%%%
\subsection{$S_2^{\Theta}$ geometry}
%%%%%%%%%%%%%%%%%%%%%%%%%%%%%%%%%%%%%

To begin with, consider the transformations, with ordinary geometries, between the cartesian ($\tau,x,y,z$) 
and spherical polar coordinates ($\tau,r,\theta,\phi$). They are given by
\bea
&&\tau\rightarrow \tau\; ,\qquad z= r \cos\theta\ ,\nonumber\\
&&x= r \sin\theta \cos\phi\;\ {\rm and}\quad y=r \cos\theta \sin\phi\ .\label{trans1}
\eea 
Let us incorporate the space-time noncommutative constraints in the ${\cal J}_C$-coordinate system to that in 
${\cal J}_S$. Since, the angle $\theta$ may be kept fixed in the 
($\tilde\tau,\tilde z$)-coordinate sector, it leads to 
\be
d\tau^2\rightarrow d{\tilde\tau}^2\;\ {\rm and}\qquad
dz^2\rightarrow d{\tilde r}^2= dr^2 \cos^2\theta\ ,\label{trans11}
\ee
where ${\tilde r}= r\cos\theta > r$, for a nontrivial $0<\theta<\pi$.
Similarly $\tilde r$ is kept fixed in the ($\tilde x,\tilde y$) coordinate sector to yield a two sphere 
with a constant radius $r_0$. Then, the relevant line element in the $L$-sector becomes
\be
(dx^2+dy^2)\rightarrow (d{\tilde x}^2+d{\tilde y}^2) = 
r_0^2\left (\cos^2\theta\ d{\theta}^2 + \sin^2\theta\ d\phi^2\right )\ .\label{trans12}
\ee
Since $\theta$ and $\phi$ are arbitrary angles in the ($\tilde x,\tilde y$)-space sector, eq.(\ref{trans12})
corresponds to a modified $S_2$ geometry. It implies that the spherical symmetry in a 
$4$ dimensional space-time is broken in a noncommutative set-up, 
$i.e.\ S_2\rightarrow S_2^{\Theta}$. The fact that $S_2^{\Theta}$ is defined 
with a constant radius $r=r_0$ is a consequence of independent $L$- and 
$\perp$-spaces, respectively, defined with small and large length scales in the theory. 
Since the above approximation becomes exact in the gravity decoupling limit \cite{kar-panda}, 
the effective description evolves with the notion of a two dimensional space-time in its semi-classical
regime \cite{kar-majumdar}. Then $r_0$ may be identified with the critical value of $\tilde r$, $i.e.\ r_0=r_c$.  
Generalization of $S_2^{\Theta}$ line element (\ref{trans12}) to that in a $4$-dimensional effective theory possibly
requires an arbitrary radial coordinate ${\tilde r}>r_c$.  
Then the relevant $4$-dimensional Euclidean line element, obtained in the
asymptotic limit, may be generalized from eqs.(\ref{trans11}) and (\ref{trans12}). In ${\tilde{\cal J}}_S$-coordinate
system, the complete line element becomes
\be
ds^2= d{\tilde\tau}^2 + d{\tilde r}^2 + {\tilde r}^2\ d\Omega^2 - {\tilde r}^2 \sin^2{\tilde\theta}\ d{\tilde\theta}^2
\ .\label{trans121}
\ee
In a re-defined orthogonal noncommutative coordinate system ${\tilde{\cal J}}_o\equiv ({\tilde\tau}, 
{\tilde r}, {\tilde l}_{\theta}, {\tilde l}_{\phi})$, the line element (\ref{trans121}) becomes
\be
ds^2= d{\tilde\tau}^2 + d{\tilde r}^2 + \cos^2\tilde\theta\ d{\tilde l}^2_{\theta} + d{\tilde l}^2_{\phi} 
\  ,\label{trans13}
\ee
\be
{\rm where}\qquad d{\tilde l}_{\theta}=\tilde r\ d\tilde\theta\qquad
{\rm and}\quad d{\tilde l}_{\phi} =\tilde r\ \sin \tilde\theta\ d\tilde\phi
\ .\qquad\qquad\label{trans2}
\ee
In addition to the correction in $S_2$ geometry, the radial coordinate is defined for large 
${\tilde r}$, $i.e.$ for ${\tilde r}>r_c$.
Then the geometry in the limit $\Theta\rightarrow 0$ may be obtained from two limiting values on coordinates in
${\tilde{\cal J}}_O$, $i.e.\ \tilde\theta\rightarrow 0$ and $r_c\rightarrow 0$. 

\sp
At this point, we recall the small radii obtained for the compact longitudinal coordinates following the 
noncommutative scaling on a $D_3$-brane \cite{kar-majumdar}. Since the quantized longitudinal space in the 
frame-work is governed by a flat metric (\ref{action9}), 
it may well be described by the $S_2^{\Theta}$ geometry in ${\tilde {\cal J}}_S$. The quantum feature claimed 
to be associated with (${\tilde\theta}, {\tilde\phi}$) coordinates is supported by the fact that the $\perp$-space 
in ${\tilde{\cal J}}_S$ is governed by large dimensions (${\tilde r}, {\tilde\tau}$). Then the $L$-space governed 
by (${\tilde l}_{\theta} ,{\tilde l}_{\phi}$) in ${\tilde{\cal J}}_O$ are, a priori, defined for
infinitesimal $\delta\tilde\theta$ and $\delta\tilde\phi$, $i.e.$ 
($0\le\tilde\theta\le\delta\tilde\theta$) and ($0\le\tilde\phi\le\delta\tilde\phi$).
In fact, the $\perp$-space may be seen to be parametrized by the plane polar coordinates 
($\tilde r, \tilde\tau$) for $\tilde r\ \epsilon (r_c, \infty)$ and $\tilde\tau\ \epsilon
[0, 4\pi \tau_c]$, where $\tau_c$ is a minimal noncommutative scale in the theory. A careful analysis in 
${\tilde{\cal J}}_o$ leads to an equivalent coordinate system with a fixed angle $\tilde\theta$ and an 
arbitrary angle $\tilde\phi\ \epsilon [0, 2\pi]$. The resulting coordinate system with  a fixed ${\tilde\theta}$ 
is in precise agreement with our approximation to obtain the line element (\ref{trans13}).  
In fact, the ${\tilde{\cal J}}_O$-system is consistent with the notion of light wedge instead of a light cone 
in a noncommutative theory \cite{alvarez-gaume}. 

%%%%%%%%%%%%%%%%%%%%%%%%%%%%%%%%%%%%%%%%%%%%%%%%%%%
\subsection{Corrections to ($\tau, r$)-coordinates}
%%%%%%%%%%%%%%%%%%%%%%%%%%%%%%%%%%%%%%%%%%%%%%%%%%%

Now we look for an appropriate transformation from the noncommutative ${\tilde{\cal J}}_S$ system 
to its ordinary counter-part in the spherical polar coordinates 
$(\tau,r,\theta, \phi )$, $i.e.$ ${\cal J}_S$-system. 
At this point, we recall the transformations obtained between the zero modes on a $D_3$-brane in ref.\cite{kar-majumdar}. 
It has been argued that the transformations incorporate (small) constant shifts on the brane coordinates and 
make them discrete.
Then the noncommutative $D_3$-brane (zero modes) coordinates are given by  
\be
{\tilde y}^{\mu} = y^{\mu} + \Theta^{\mu\nu} p_{\nu}\ ,\label{black01}
\ee
where $p_{\mu}$ is the canonical conjugate momentum to $y^{\mu}$.
Explicitly, the shift transformations are worked out, in ${\tilde{\cal J}}_C$-system, to yield
\bea
\tilde\tau &=&\tau + \big ({\bf\Theta}_E\cdot {\bf p}\big ) \nonumber\\
{\rm and}\quad {\tilde r}^i &=& r^i + \epsilon^{ijk} \Theta_B^k\ p_j + \Theta_E^i\ p_4\ ,\label{black025}
\eea
where $\Theta^{4i}\equiv\Theta_E^i= (0, \Theta_E^y, \Theta_E^z)$ and 
$\Theta_B^i= (0, \Theta_B^y, \Theta_B^z)$. Then ${\tilde r}^2$ becomes
\be
{\tilde r}^2 = r^2 + r^2_c\ .\label{black02}
\ee
Explicitly, $r^2_c$ is given by
\bea
r^2_c&=&\big ({\bf\Theta}_B\cdot {\bf L}\big )\ +\ 
\big ({\bf\Theta}_E\cdot {\bf r}\big )p_4 \ +\ {\cal O}(\Theta^2)
\ +\ \dots \ ,\nonumber\\
&\equiv& \Theta + {\cal O}(\Theta^2)\ +\ \dots\ ,\label{black022}
\eea
where ${\bf L}$ denotes the angular momentum vector and $|{\bf \Theta_B}|=|{\bf \Theta_E}|$.
It shows that the every coordinate in the corresponding cartesian system ${\tilde{\cal J}}_C$ are bounded from below. 
In fact the bound on the radial coordinate, $i.e.\ r_c$, arises from the upper bound on the ${\bf E}$ field in the theory.

\sp
On the other hand, the transformation between ${\tilde{\cal J}}_C\rightarrow {\tilde{\cal J}}_O$ may 
equivalently be viewed as an orthogonal transformation between its commutative counter-parts. In particular, 
it is given by an orthogonal rotation ${\cal R}_{\theta}$ of ($r, \theta$)-coordinates by an arbitrary polar angle $\theta$ 
around the symmetry axes ($\tau$, $\phi$). The transformations may be given by 
\be
{\cal J}_C\rightarrow {\cal J}_S\equiv {\tilde{\cal J}}_O + {\cal R}_{\theta}\ .\label{rot}
\ee
It implies that there are several coordinate systems equivalent to ${\tilde{\cal J}}_O$ and 
they are defined for different values $\theta$ in the range $\theta\ \eps (0,\pi)$. 
In other words, the continuous coordinate $\theta$ in the $S_2$ geometry, picks up 
discrete values in its noncommutative counter-part $S_2^{\Theta}$. As a result the spherical symmetry 
around an axis is broken in the theory. 
However the configuration space involving all possible rotations ${\cal R}_{\theta}$, 
preserves the spherical symmetry in the noncommutative frame-work. Thus the inclusion of configuration 
space for $\theta$, in ${\tilde{\cal J}}_O$, results in a spherically symmetric coordinate system equivalent 
to ${\cal J}_S$. 

%%%%%%%%%%%%%%%%%%%%%%%%%%%%%%%%%
\section{Generalized black holes}
%%%%%%%%%%%%%%%%%%%%%%%%%%%%%%%%%

%%%%%%%%%%%%%%%%%%%%%%%%%%%%%%%%%%%%%%%%%%
\subsection{Schwarzschild like geometries}
%%%%%%%%%%%%%%%%%%%%%%%%%%%%%%%%%%%%%%%%%%

In this section, we construct a generalized black hole geometry in the effective noncommutative theory (\ref{action7}).
In particular, we attempt to formulate the Schwarzschild like geometry (\ref{act10}) obtained, with an ordinary space-time,
to the noncommutative one. Since the effective metric (\ref{act33}) is used to obtain the Schwarzschild like black hole, the
underlying theory is naturally governed by the noncommutative formalism at Planck energy. 

\sp
To begin with, we rather focus on the ordinary counter-part (\ref{action116}) of the noncommutative string theory 
(\ref{action42}). The self-duality in the gauge field yields $Q_e=Q_m=Q$. 
However the energy momentum tensor $T_{\mu\nu}=0$ in the theory. Thus the relevant dynamics is essentially described 
by the vacuum Einstein's equations of motion. Its solution is given by the Schwarzschild black hole.
In addition, the gauge potential (\ref{act81}) in the theory consistently describes a parallel field 
configuration, $i.e.\ {\bf E} = \big ( E_{\hat r},0,0\big )$ and ${\bf B}= \big (B_{\hat r},0,0\big )$. The
EM-field is obtained (\ref{act82}) may be be checked to be in agreement with the duality invariance 
(\ref{action114}) in the theory.

\sp
Now a generalized semi-classical solution on a curved $D_3$-brane (\ref{action4}), 
may be constructed using the  effective metric (\ref{act33}) in the theory. Interestingly for ${\cal B}=0$, the
semi-classical solution has been obtained in eq.(\ref{act10}). However, in presence of ${\cal B}$-field, the
effective metric in ${\tilde{\cal J}}_O$-system may be expressed as
\bea
ds^2&=&\left (1-{{2G_NM}\over{\tilde r}} - {{(G_NQ)^2}\over{{\tilde r}^4}} + 
{{2G_N^3MQ^2}\over{{\tilde r}^5}}\right ) d{\tilde\tau}^2 \nonumber\\
&&\qquad + \left (1- {{2G_NM}\over{{\tilde r}}} - {{(G_NQ)^2}\over{{\tilde r}^4}} + 
{{2G_N^3MQ^2}\over{{\tilde r}^5}}\right )^{-1} d{\tilde r}^2 \nonumber\\
&&\qquad\qquad 
+\ \left ( 1 - {{(G_NQ)^2}\over{{\tilde r}^4}}\right ) d{\tilde l}^2_{\theta} 
+\ \left ( 1 - {{(G_NQ)^2}\over{{\tilde r}^4}}\right )^{-1} d{\tilde l}^2_{\phi}
\ ,\label{black5b}
\eea
The generalized solution breaks the spherical symmetry, which is due to the fact that $\tilde\theta$ is fixed in 
${\tilde{\cal J}}_O$-coordinate system. 
The gauge potential ${\hat A}_{\mu}$ is worked out, from eq.(\ref{black211}) using eq.(\ref{act81}), 
in the theory. Since the constraints (\ref{action9}) ensure ${\hat A}_{\tilde\phi}=0$, the 
generalized solution (\ref{black5b}) may be seen to be governed by a trivial noncommutative gauge potential
${\hat A}_{\mu} = 0$.

\sp
The curvature singularity in the generalized solution (\ref{black5b}) appears 
at ${\tilde r}=r_c$. The solution possesses a coordinate singularity at $r_c=\pm {\sqrt{G_NQ}}$ 
in addition to its event horizon at $r_s=2G_NM$, where $r_s>r_c$. Since $\tilde r$ possesses a lower bound at
${\tilde r}=r_c$, the curvature singularity in the solution (\ref{black5b}) is governed only at its lower bound.
In other words, the solution (\ref{black5b}) describes a GS-black hole geometry in the
effective theory. Though the GS-solution possesses both (equal) electric and magnetic charges, 
its event horizon is actually governed by its Schwarzschild radius $r_s$. This is due to the underlying fact that
$T_{\mu\nu}=0$ in the theory. On the other hand, the exact nature (in $\Theta$) of the effective theory makes it
accessible to its electric charge $Q$ at high energy. Thus $Q$ adds a small mass and may be seen to be 
associated with the angular momentum (\ref{black022}) in the effective noncommutative theory.  

\sp
The GS-black hole (\ref{black5b}) on a noncommutative $D_3$-brane, essentially describes 
an Euclidean manifold $R^2\times S^2$ and is valid for large (${\tilde r}, {\tilde\tau}$), 
$i.e.$ for ${\tilde r}\ge r_c$ and ${\tilde\tau}\ge \tau_c$.
Since the $\perp$-space is spanned by the large dimensions
in a noncommutative frame-work \cite{kar-panda,kar-majumdar}, they may be well described by  
($\tilde r,\tilde\tau$) in the frame-work. 
The remaining two orthogonal coordinates ($l_{\tilde\theta}$, $l_{\tilde\phi}$) there describe 
the quantized $L$-plane. 

\sp
\noindent
Using eq.(\ref{black02}), the relations between the parameters in the two coordinate systems 
${\tilde{\cal J}}_O$ and ${\cal J}_S$ are worked out to yield
\bea
&&\quad M_{\rm eff}= (G_NM)\left [ 1 - {{\Theta}\over{2r^2}} +\ {\cal O}(\Theta^2)\ +\ \dots\ \right ]
\qquad\qquad\qquad\qquad\nonumber\\
{\rm and}&&\quad {r^2_c}= (G_N Q)
\left [1 - {{\Theta}\over{r^2}} +\ {\cal O}(\Theta^2)\ +\ \dots\ \right ]\ .\label{black52}
\eea
Finally, the generalized geometry (\ref{black5b}), in ${\cal J}_S$-coordinates, becomes
\bea
ds^2&=& \left (1-{{2M_{\rm eff}}\over{r}} - {{r_c^4}\over{r^4}} + {{2M_{\rm eff}\ r_c^4}\over{r^5}}\right ) d\tau^2
+ \left (1-{{2M_{\rm eff}}\over{r}} - {{r_c^4}\over{r^4}} + {{2M_{\rm eff}\ 
r_c^4}\over{r^5}} \right )^{-1} dr^2\nonumber\\
&&\quad\qquad\qquad\qquad\qquad\qquad + \left ( \ 1 - {{r_c^4}\over{r^4}} \right )
r^2 d\theta^2 \ +\ \left ( \ 1 - {{r_c^4}\over{r^4}} \right )^{-1} 
r^2 \sin^2 \theta d\phi^2 \ .\label{black6b} 
\eea
The curvature singularity in the GS-solution occurs at $r=0$ in ${\cal J}_S$-system. 
In addition, the geometry possesses two distinct coordinate singularities at $r= r_c$ and $r_s$. 
Since $r_c<r_s$, the curvature singularity of the semi-classical black hole 
is well protected by its event horizon at $r_c$ in the frame-work. The other coordinate singularity at $r_s$
is in the classical regime. It may not serve the purpose of an event horizon and hence we 
call it a ``pseudo horizon'' in the theory. In fact, the GS-geometry at its Schwarzschild radius possesses
all properties of that of an event horizon for $M\neq 0$. The possible cause of hindrance at $r_s$ to 
treat it as an event horizon may be resolved by the fact that the $D_3$-brane is flat in the classical regime, 
$i.e.\ M=0$ on its world-volume. Then an observer's motion on a $D_3$-brane world-volume 
appears to be ``tangential'' to the event horizon of the classical GS-black hole. However the inherent
noncommutative scale in the theory deform the tangential touch of the horizon to a finite small cut on the $D_3$-brane.
Heuristically, the observer is rather just inside or just outside the event horizon $r_s$ of a typical 
Schwarzschild black hole.
At this point, we recall an $e^-e^+$ pair production process, just outside $r_s$, initiated by the uniform
${\bf E}$-field on the $D_3$-brane in the classical regime. If an $e^-$ moves in the out-ward direction, then
the $e^+$ would traverse the coordinate singularity at $r_s$ in the in-ward direction. Subsequently 
the $e^+$ behaves like an $e^-$ inside $r_s$ and its energy becomes negative, which is required by the energy 
conservation in the pair production process. Since the event horizon on the
$D_3$-brane is actually governed at $r_c$, the observer would experience an effective gravitational force due to the 
${\bf E}$-field in the classical regime. 

\sp
Further more, the pair creation phenomenon by an uniform ${\bf E}$-field may be seen to explain the Hawking radiation 
\cite{hawking1} from the generalized GS-geometry (\ref{black6b}). In the context, we find that eq.(\ref{black52}) shows 
the presence of effective parameters, $M_{\rm eff}$ and $r_c$ in the generalized black hole in a noncommutative formalism. 
Since $\Theta$-terms modify the mass of the black hole, its Schwarzschild radius  $r_s=2M_{\rm eff}$  is not fixed in 
the ${\cal J}_S$-system. It confirms that the semi-classical GS-black hole Hawking radiates. 
The source of radiation is primarily due to the $\Theta$-terms, which give rise to the  nonlinearity in the 
${\bf E}$-field. Since the nonlinearity is determined by the non-vanishing global mode of the ${\bf E}$-field, the
source of radiation is uniform in the frame-work. An $e^-e^+$ pair created by an uniform ${\bf E}$-field at the
vacuum solution to the Einstein's equations of motion may be viewed as that produced just outside $r_s$. As pointed out, 
an $e^-$ motion in the out-ward direction from $r_s$ would associate the $e^+$ motion in the in-ward direction. 
Since the velocity of a particle is zero at $r_s$, the motion of the negative energy $e^+$ may be viewed as that of a 
positive energy (secondary) $e^-$ in the out-ward direction.  However, the secondary $e^-$ is time dilated as 
it overcomes the horizon from inside. The pair production continues with the increase of ${\bf E}$-field, at each step, 
until ${\bf E}_C$ is reached. The $e^+$ from the second pair produced at $r_s$ is attracted by the dilated $e^-$, they 
annihilate to produce the quantum of radiation in the theory. In each step, the event horizon at $r_s$ shrinks in its size. 
The Hawking temperature for the GS-black hole is worked out to yield
\be
T^{\rm GS}_{\rm Hawking}={1\over{8\pi M_{\rm eff}}}\ .\label{hawk1}
\ee
It implies that the temperature increases in each step, followed by the quantum radiation, in the theory. 

\sp
On the other hand, the higher order corrections, in $1/r$ to the Einstein's metric dominate at the Planck scale, 
$i.e.$ in the limit ${\bf E}\rightarrow {\bf E}_c$. They are indeed associated with 
the electric charge in the theory. The correction term, $i.e.\ Q^2/r^4$, to the Einstein's metric $g_{\mu\nu}$ 
may be identified with the energy momentum tensor computed in a semi-classical theory governed by the Einstein's 
theory coupled to the Maxwell's.
In other words, they contribute a mass correction (\ref{black52}) to the Schwarzschild mass $M$ otherwise
present in the theory.
At Planck scale, since $Q^2/r^4$ dominates over $2M_{\rm eff}/r$, the $4$-dimensional 
generalized geometry (\ref{black6b}) may alternately be viewed as a higher dimensional, semi-classical, Schwarzschild 
black hole produced in the scattering of noncommutative strings \cite{kar-majumdar}. In particular an $d$-dimensional 
Schwarzschild black hole mass is known to be associated with $r^{(3-d)}$ term in the metric. Then 
the semi-classical black hole produced in a high energy collision may naively be argued to have its origin in 
$7$-dimensions. However the noncommutative constraints in the theory reduce the space-time dimensions
to five from seven. Interestingly, the effective  $4$-dimensional space-time (\ref{black6b}) may be obtained from a 
$5$-dimensional gravity ($eg.$ see \cite{kmp,eardley-giddings}) with a small scale along the fifth dimension. 
The generalized black hole (\ref{black6b}) in an effective theory of gravity 
is identical to that obtained in the Einstein's GTR. However in the classical regime,
the corrections in the  effective metric are small and may be ignored.
Then, the generalized geometry (\ref{black6b}) precisely reduces to the Schwarzschild black hole. 

%%%%%%%%%%%%%%%%%%%%%%%%%%%%%%%%%%%%%%%%%%%%%%%
\subsection{Reissner-Nordstrom like geometries}
%%%%%%%%%%%%%%%%%%%%%%%%%%%%%%%%%%%%%%%%%%%%%%%

Let us recall the approximations enforced on the ${\tilde{\cal J}}_S$-coordinates, which in turn lead to a more
appropriate ${\tilde{\cal J}}_O$-system on the noncommutative $D_3$-brane.
Since $\tilde\theta\rightarrow 0$ in ${\tilde{\cal J}}_O$-system, an equivalent arbitrary $\theta$ description 
(\ref{rot}) in ${\cal J}_S$-coordinates give rise to useful insights into the theory.

\sp
Now let us consider the the general solution  (\ref{action9}) in ${\tilde{\cal J}}_O$-system.
The potential in its general form becomes
\be
{\hat A}_{\mu} = \left ({\hat A}_{\tilde\tau}, {\hat A}_{\tilde r},0,0 \right ) \ .\label{black22}
\ee
Eq.(\ref{action10}) confirms that the components of gauge field in ${\tilde{\cal J}}_O$ coordinates satisfy 
\bea 
&&\left ( \partial^2_{{\tilde l}_{\theta}} + \partial^2_{{\tilde l}_{\phi}}\right ) {\hat A}_{\tilde\tau} = 0
\nonumber\\
{\rm and}\quad
&&\left ( \partial^2_{{\tilde l}_{\theta}} + \partial^2_{{\tilde l}_{\phi}}\right ) {\hat A}_{\tilde r} = 0\ .
\qquad\qquad\qquad\qquad\label{black221}
\eea
Then the noncommutative gauge potential becomes trivial in ${\tilde{\cal J}}_O$ coordinates. However
in ${\tilde{\cal J}}_S$-coordinates, $i.e.$ for arbitrary $\tilde\theta$, the non-vanishing orthogonal 
components of the potential are given by
\be
{\hat A}_{\tilde\tau}= -i{\hat Q}\ \sin \tilde\theta \cos \tilde\phi\quad {\rm and}\qquad 
{\hat A}_{\tilde r}= {\hat Q}\ \sin \tilde\theta \sin \tilde\phi
\ ,\label{black222}
\ee
where ${\hat Q}$ is a constant and  may be interpreted as an electric charge in the noncommutative theory. 
Then the orthogonal components of the EM-field, in ${\tilde{\cal J}}_S$-system, are computed to yield
\bea
&&E_{\hat\theta}= {{\partial_{\theta}{\hat A}_{0}}\over{\tilde r \cos \tilde\theta}}
= {{\hat Q}\over{\tilde r}}\cos \tilde\phi\ ,\nonumber\\
&&E_{\hat\phi}= {{\partial_{\phi}{\hat A}_{0}}\over{\tilde r \sin \tilde\theta}}
= -{{\hat Q}\over{\tilde r}}\sin \tilde\phi\nonumber\\
{\rm and }\; &&B_{\hat\theta} = {{\partial_{\hat\phi}{\hat A}_{\hat r}}\over{\tilde r \sin \tilde\theta}}
= {{\hat Q}\over{\tilde r}}\cos \tilde\phi\ ,\nonumber\\
&&B_{\hat\phi} = - {{\partial_{\hat\theta}{\hat A}_{\hat r}}\over{\tilde r \cos \tilde\theta}}
= - {{\hat Q}\over{\tilde r}}\sin \tilde\phi\ .\label{rnem}
\eea
The ${\bf E}$- and ${\bf B}$-field components are satisfy the duality invariance (\ref{action114}) 
in the theory. In fact, they (\ref{rnem}) describe a parallel EM-field in ${\tilde{\cal J}}_O$-coordinates.
The energy momentum tensor (\ref{action12}) in the theory is computed to yield 
\bea
{\hat T}_{\tau\tau}&=-&{{i{\hat Q}^3\Theta_E}\over{{\tilde r}^3}}\left [ \cos \tilde\phi -\sin \tilde\phi \right ]\nonumber\\
&=&-{{i{\hat Q}^2}\over{{\tilde r}^2}}\left [ \cos \tilde\phi -\sin \tilde\phi \right ]\ .\label{emt1} 
\eea
It implies that the limit $\Theta_E\rightarrow 0$, for an arbitrary angle $\tilde\phi$, may alternately
be described by $\phi=\pi/4$ for a fixed $\Theta$. 
On the other hand, the  ${\bf E}$- and ${\bf B}$-fields (\ref{rnem}) may be incorporated into the modified metric 
(\ref{act33}) for a vacuum solution to Einstein's theory. 
In the ${\tilde{\cal J}}_O$-coordinates, the effective solution becomes
\bea
ds^2&=&\left (1-{{2G_NM}\over{\tilde r}} - {{G_N{\hat Q}^2}\over{{\tilde r}^2}} \right ) d{\tilde\tau}^2 
+\left (1- {{2G_NM}\over{\tilde r}} - {{G_N{\hat Q}^2}\over{{\tilde r}^2}} \right )^{-1} d{\tilde r}^2 
\qquad\qquad\nonumber\\
&&\;\ \qquad\qquad\qquad\qquad\qquad\qquad 
+\left (1 -\ {{G_N{\hat Q}^2}\over{{\tilde r}^2}} \cos 2\tilde\phi - {{2G_N^2M{\hat Q}^2}\over{{\tilde r}^3}} \right ) 
d{\tilde l}^2_{\theta}\nonumber\\
&&\;\ \qquad\qquad\qquad\qquad\qquad\qquad +\ \left (1 +\ {{G_N{\hat Q}^2}\over{{\tilde r}^2}} \cos 2\tilde\phi - 
{{2G_N^2M{\hat Q}^2}\over{{\tilde r}^3}} \right ) d{\tilde l}^2_{\phi}\ .\label{black23}
\eea
The generalized geometry possesses a curvature singularity at ${\tilde r}={\hat Q}$.
The solution describes a GRN-black hole in a noncommutative open string theory with event horizons at 
\be
r_{\pm}= (G_NM) \pm {\sqrt{(G_NM)^2 - G_N{\hat Q}^2}}\ .\label{black24}
\ee
Thus the noncommutative $D_3$-brane world-volume geometry may also be governed by a 
GRN-black hole (\ref{black23}), in addition to the GS-geometry (\ref{black6b}). 
Using eq.(\ref{black02}), we obtain 
\bea
{\hat Q}^2_{\rm eff}&=&(G_N{\hat Q}^2)\left [ 1 -{{\Theta}\over{r^2}}\ +\ 
{\cal O}(\Theta^2)\ +\ \dots \ \right ]\nonumber\\
&=& (G_N{\hat Q}^2) - (G_N{\hat Q}^2_{\Theta})\left [ 1 -{{\Theta}\over{r^2}}\ +\ 
{\cal O}(\Theta^2)\ +\ \dots \ \right ]\ ,\qquad\qquad\qquad\qquad 
\label{black26}
\eea
where ${\hat Q}_{\Theta}$ is a non-zero constant in the theory.   
Then the GRN-black hole (\ref{black23}), in ($\tau, r, \theta, \phi$) coordinates may be re-expressed as
\bea
ds^2&=&\left (1-{{2M_{\rm eff}}\over{r}} \pm {{{\hat Q}^2_{\rm eff}}\over{r^2}} \right ) d\tau^2 
+\left (1- {{2M_{\rm eff}}\over{r}} \pm {{{\hat Q}^2_{\rm eff}}\over{r^2}} \right )^{-1} dr^2 
\qquad\nonumber\\
&&\qquad\qquad\qquad\qquad\ +\ 
\left (1 \pm\ {{{\hat Q}^2_{\rm eff}}\over{r}^2} \cos 2\phi \pm
{{2M_{\rm eff}{\hat Q}^2_{\rm eff}}\over{r^3}} \right ) 
r^2 d\theta^2\nonumber\\
&&\qquad\qquad\qquad\qquad\ +\ \left (1 \mp\ {{{\hat Q}^2_{\rm eff}}\over{r^2}} \cos 2\phi \pm
{{2M_{\rm eff}{\hat Q}^2_{\rm eff}}\over{r^3}} \right ) r^2 \sin^2\theta\ d\phi^2\ .\label{black27}
\eea
As expected in a noncommutative theory, the black hole geometry is not spherically symmetric. It
reconfirms that for $\phi=\pi/4$, the $S_2$ symmetry is restored which is equivalently described by the 
$\Theta\rightarrow 0$. 

\sp
\sp
\noindent
{\bf Semi-classical GRN-geometry}

\sp
The GRN-solution (\ref{black27}) in the regime is described by the $(+)$ve sign in its metric components.
The semi-classical GRN-solution may be seen to possess a curvature singularity at $r=0$. 
It describes a charged black hole with event horizons at 
\be
r_{\pm}=M_{\rm eff} \pm {\sqrt{M^2_{\rm eff} - {\hat Q}^2_{\rm eff}}}\ .\label{black28}
\ee
Since the mass $M_{\rm eff}$ and charge ${\hat Q}_{\rm eff}$ 
are not constants, the semi-classical GRN-black hole exhibits Hawking radiation.
The Hawking temperature may be computed to yield
\be
T^{\rm GRN}_{\rm Hawking} = {{\sqrt{M^2_{\rm eff} - {\hat Q}^2_{\rm eff}}}\over{2\pi\left [ M_{\rm eff} 
+ {\sqrt{M^2_{\rm eff} - {\hat Q}^2_{\rm eff}}}\right ]^2}}\ .\label{hawk2}
\ee
In the limit $\Theta\rightarrow 0$, the GRN-geometry in the regime 
corresponds to the Reissner-Nordstrom (RN-) like geometry. It is given by
\bea
ds^2&=&\left (1-{{2M}\over{r}} + {{G_N{\hat Q}^2}\over{r^2}} \right ) d\tau^2 
+\left (1 + {{2M}\over{r}} + {{G_N{\hat Q}^2}\over{r^2}} \right )^{-1} dr^2 \qquad\nonumber\\
&&\qquad +\ 
\left (1 +\ {{G_N{\hat Q}^2}\over{r}^2} \cos 2\phi \right ) r^2 d\theta^2
\ +\ \left (1 -\ {{G_N{\hat Q}^2}\over{r^2}} \cos 2\phi 
\right ) r^2 \sin^2\theta\ d\phi^2\ .\label{black272}
\eea
In the limit $\phi=\pi/4$, then the solution precisely describes the 
RN-black hole goemetry. It may be seen to exhibit Hawking radiation in the frame-work. The Hawking temperature 
is given by
\be
T^{\rm RN}_{\rm Hawking} = {{\sqrt{{M^2} - G_N^{_1}{\hat Q}^2}}\over{2\pi\left [ M 
+ {\sqrt{M^2 - G_N^{-1}{\hat Q}^2}}\right ]^2}}\ .\label{hawk22}
\ee
The temperature decreases with the quantum radiations from the semi-classical RN-black hole. 
With the emission of radiation, $M$ decreases and finally becomes equal to its charge ${\hat Q}/{\sqrt{G_N}}$.
The process ceases as $T^{\rm GRN}_{\rm Hawking}\rightarrow 0$ in the limit $M\rightarrow {\hat Q}/{\sqrt{G_N}}$, 
which may be identified with the gravity decoupling limit in the theory. 
In other words, in the limit of vanishing nonlinearity ${\cal B}=0$, the GRN-geometry 
in the classical regime reduces to a typical RN-black hole, which is described by the linear ${\bf E}$-field.
Schematically,
the noncommutative $D_3$-brane geometries in ${\tilde{\cal J}}_O$-system for various values of $r$ 
is illustrated in fig.2.
%\ref{sksmf2}
\begin{figure}[ht]
\begin{center}
\vspace*{3.5in}
%\hspace*{-.5in}
\relax{\includegraphics{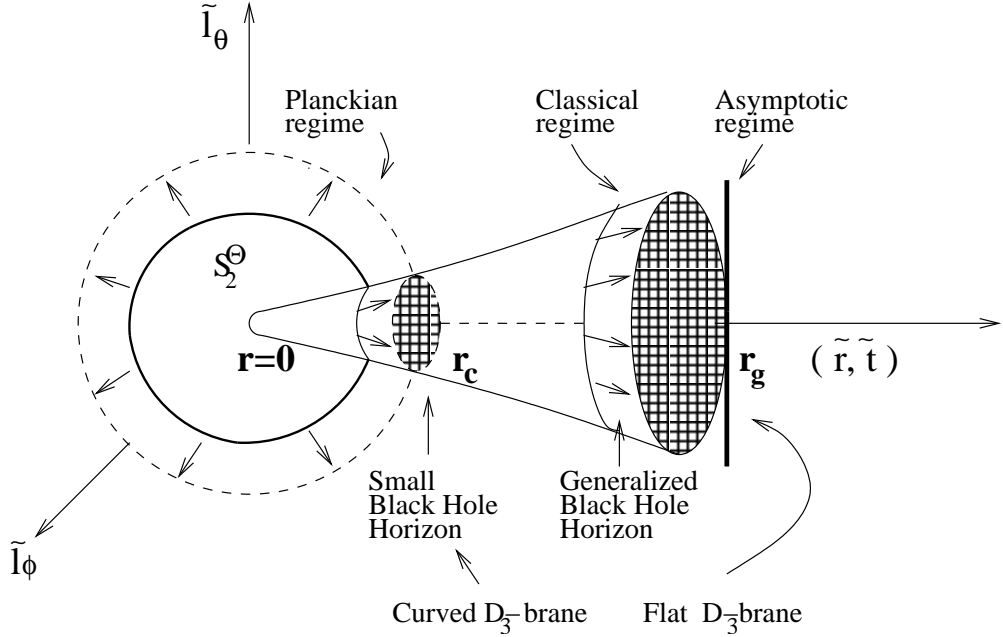}}
\end{center}
\vspace*{-.1in}
\caption{Noncommutative $D_3$-brane (sphere) geometries primarily at two different length scales $l_s$ and $l_{\perp}$.
They are characterized by Planckain ($r\rightarrow r_c$), classical ($r_c<r\le r_g$) and asymptotic ($r_g\le r\le\infty$)
regimes. The curved $D_3$-brane is defined for $r\le r_g$ and the flat brane is defined in the asymptotic regime there.}
\label{sksmf2}
\end{figure}

\sp
\noindent
{\bf GRN-geometry at Planck scale}

\sp
The $(-)$ve sign in the GRN-geometry (\ref{black27}) leads to an apparent contradiction to the 
nature of a Reissner-Nordstrom (RN-) like black hole. The Hawking temperature in the regime becomes
\be
T^{\rm GRN}_{\rm Hawking} = {{\sqrt{M^2_{\rm eff} + {\hat Q}^2_{\rm eff}}}\over{2\pi\left [ M_{\rm eff} 
+ {\sqrt{M^2_{\rm eff} + {\hat Q}^2_{\rm eff}}}\right ]^2}}\ .\label{hawk3}
\ee
The temperature increases with the emission of 
quantum radiations from the GRN-black hole, which is unlike to that of a typical semi-classical RN-black hole.
However, the contradiction may be resolved by the fact that the noncommutative theory is scale dependent. 
In the regime, the GRN-solution is governed by a GS-geometry, which is evident from the increase in 
Hawking temperature (\ref{hawk3}) in the theory. The phenomenon is supported by the fact that the GRN-black hole
becomes an extremum in the regime. GS-geometry becomes obvious when two black holes (\ref{black6b}) and 
(\ref{black27}) are compared 
with respect to their metric components. Since the higher order terms in $1/r$ become significant in 
the regime, the GRN-geometry
is effectively governed by the GS-black hole. For instance a leading order term, in $\Theta$, present in the GS-geometry 
may be seen to arise from the linear order term in the GRN-black hole and so on. 

%%%%%%%%%%%%%%%%%%%%%%%%%%%%%%%%%%%%%
\section{Two dimensional black holes} 
%%%%%%%%%%%%%%%%%%%%%%%%%%%%%%%%%%%%%

%%%%%%%%%%%%%%%%%%%%%%%%%%%%%%%%%%%%
\subsection{Semi-classical geometry}
%%%%%%%%%%%%%%%%%%%%%%%%%%%%%%%%%%%%

Now let us focus on the near horizon geometry in the semi-classical regime obtained from the 
GRN-black hole (\ref{black27}). Explicitly it is given by
\bea
ds^2&=&\left (1-{{2M_{\rm eff}}\over{r}} + {{{\hat Q}^2_{\rm eff}}\over{r^2}} \right ) d\tau^2 
+\left (1- {{2M_{\rm eff}}\over{r}} + {{{\hat Q}^2_{\rm eff}}\over{r^2}} \right )^{-1} dr^2 
\qquad\nonumber\\
&&\qquad\qquad\qquad\qquad\ +\ 
\left (1 +\ {{{\hat Q}^2_{\rm eff}}\over{r}^2} \cos 2\phi +
{{2M_{\rm eff}{\hat Q}^2_{\rm eff}}\over{r^3}} \right ) 
r^2 d\theta^2\nonumber\\
&&\qquad\qquad\qquad\qquad\ +\ \left (1 -\ {{{\hat Q}^2_{\rm eff}}\over{r^2}} \cos 2\phi +
{{2M_{\rm eff}{\hat Q}^2_{\rm eff}}\over{r^3}} \right ) r^2 \sin^2\theta\ d\phi^2\ .\label{scrn}
\eea
The $\perp$-space may be seen to be redefined by a set of polar coordinates ($\rho,\omega$) 
for $0\le \rho\le\infty$ and  $0\le\omega\le 2\pi$. We define 
\be
\left ( 1 - {{2M_{\rm eff}}\over{r}}\right ) = {{{\rho}^2}\over{8M_{\rm eff}}}\ ,\label{2d1}
\ee
then in the limit $r\rightarrow r_s (=2M_{\rm eff})$, the radial coordinate becomes $\rho\rightarrow 0$. Since 
${\hat Q}^2_{\rm eff}/r^2\rightarrow 0$ in the limit, the geometry reduces to that of a GS-black hole. Then, the 
semi-classical GRN-black hole (\ref{black27}) is simplified for its near horizon geometry to yield an Euclidean 
Rindler space-time in the $\perp$-plane. On the other hand, the $L$-space there, describes the high energy modes 
and decouples completely in the limit $r\rightarrow r_s$. The near horizon geometry in GRN-black hole in the
semi-classical regime may be approximated by
\be
ds^2= d{\rho}^2 + {{4{\rho}^2}\over{r_s^2}} d\tau^2 + 8r_s^2 \ d\Omega^2
\ .\label{2d2}
\ee
It shows that the $\perp$- and $L$-spaces are completely independent in the near horizon geometry of the
semi-classical GRN-black hole. In the regime, the effective space-time is governed by the $\perp$-space is 
given by
\be
ds^2_{\perp}= d{\rho}^2 + {\rho}^2 d{\omega}^2 \ , \label{2d3}
\ee 
where $\omega= 2\tau/r_s$. Then, the Euclidean time becomes $\tau\rightarrow \tau + 2 \pi M_{\rm eff}$.
It implies that the radius of the Euclidean time like coordinate is large in the near horizon geometry. 
On the other hand, the $L$-space in the effective $4$-dimensional frame-work is described by the
$S_2$ geometry. Thus the relevant near horizon semi-classical GRN geometry on a noncommutative $D_3$-brane 
is governed by a two dimensional black hole (\ref{2d3}) along its $\perp$-space.  

\sp
With an appropriate re-parametrization, $i.e.\ \rho= \sinh r$, the semi-classical black hole solution 
(\ref{2d3}) may be seen to describe Witten's $2D$ black hole \cite{witten91}
\be
ds^2_{\perp}= dr^2 - \tanh^2 r\ dt^2\ . \label{2d4}
\ee 
The black hole geometry is known to be governed by an exact conformal field theory. The event horizon of the
$2D$-black hole (\ref{2d4}) is at $r=0$. The scalar curvature, $R_h=4/\cosh^2 r$, remains 
regular at the event horizon.

%%%%%%%%%%%%%%%%%%%%%%%%%%%%%%%%%%%%%
\subsection{Gravity decoupled regime}
%%%%%%%%%%%%%%%%%%%%%%%%%%%%%%%%%%%%%

The gravity decoupling limit in the effective noncommutative theory (\ref{action7}) is
described by the decoupling of all the closed string modes in the theory. The limit may be incorporated
into the theory by simply taking $M\rightarrow 0$, which is equivalently represented by $M\rightarrow 
{\hat Q}/{\sqrt{G_N}}$. In other words, the limit leads to the Planckian regime in the frame-work.
It is interesting to see that the GRN-geometry (\ref{black27}) in the limit reduces to an extremum 
GRN-black hole, which in turn may be identified with the GS-black hole solution (\ref{black6b}). In other words,
the $D_3$-brane geometry in the Planckian regime is governed uniquely by the Schwarzschild like black hole.

\sp
To begin with, we consider the GS-black hole (\ref{black6b}) in the 
gravity decoupling limit, $i.e\ r\rightarrow r_c$. Then the reduced geometry may be approximated by 
\be
ds^2=\left (1- {{r^4_c}\over{r^4}} \right ) \left [d\tau^2 +r^2 d\theta^2\right ]
+\left (1 + {{r^4_c}\over{r^4}} \right ) \left [dr^2 + r^2 \sin^2\theta d\phi^2\right ]
\ .\label{sl1}
\ee
It describes a $4$-dimensional semi-classical Schwarzschild black hole in a gravity decoupled 
noncommutative string theory. 
The black hole (\ref{sl1}) possesses a small mass $r_c$ and its event horizon is at $r=r_c$. 
It describes a laboratory black hole produced in an intermediate state of the Planckian energy 
scattering phenomenon investigated in ref.\cite{kar-majumdar}.{\footnote{Interestingly, 
these laboratory black holes production 
have drawn considerable attention in the low scale effective gravity 
models, $eg.$ see \cite{emparan}-\cite{kar-jain-panda}.}} 

\sp
On the other hand, the GRN-black hole (\ref{black27}), in the 
gravity decoupling limit $r\rightarrow {\hat Q}_{\rm eff}$ may be worked out to yield 
\bea
&&ds^2=\left (1- {{{\hat Q}^2_{\rm eff}}\over{r^2}} \right ) d\tau^2 
+ \left (1 + {{{\hat Q}^2_{\rm eff}}\over{r^2}} \right ) dr^2 + \left ( 1 - {{{\hat Q}^2_{\rm eff}}\over{r^2}}
\cos 2\phi\ \right ) r^2 d\theta^2\qquad\qquad\nonumber\\
&&\qquad\qquad\qquad\qquad\qquad\qquad\qquad\qquad\qquad + 
\left ( 1 + {{{\hat Q}^2_{\rm eff}}\over{r^2}}\cos 2\phi\ \right ) r^2 \sin^2\theta\ d\phi^2
\ .\label{sl2}
\eea
The geometry describes a $4$-dimensional Euclidean 
black hole with event horizon at $r=\pm {\hat Q}_{\rm eff}$. The black hole possesses a small
mass due to the nonlinear electromagnetic field and may be seen to describe the laboratory black hole (\ref{sl1}). 

\sp
In the regime, the near horizon geometry for the laboratory black hole (\ref{sl2}) 
may be worked out in a similar way to that for the semi-classical black hole in the previous section.
The $L$-plane may appropriately be defined by the plane polar coordinates
$(\hat\rho,\hat\omega)$, where
\be
\left (1- {{{\hat Q}^2_{\rm eff}}\over{r}}\right ) = {{{\hat\rho}^2}\over{4{\hat Q}_{\rm eff}}}\ .\label{sl3}
\ee
Then, the near horizon limit is described by a new coordinate $\hat\rho$ for its small value and by $\hat\omega$ for 
$0\le \hat\omega\le 2\pi$. In the limit $r\rightarrow {\hat Q}_{\rm eff}$, 
the laboratory black hole (\ref{sl1}) is simplified using eq.(\ref{sl2}). The near horizon geometry becomes
\be
ds^2=d\rho^2 + {4{\rho^2}\over{{\hat Q}^2_{\rm eff}}}d\tau^2 + 8{\hat Q}^2_{\rm eff}\left [(1- \cos 2\phi) d\theta^2 
+ (1+ \cos 2\phi) \sin^2\theta d\phi^2\right ]
\ .\label{sl4}
\ee
In the regime, the relevant geometry reduces to
\be
ds^2_{L}= {d{\hat\rho}^2} + {\hat\rho}^2 d{\hat\omega}^2 \ , \label{sl5}
\ee 
where ${\hat\omega}= 2\tau/{\hat Q}_{\rm eff}$. As a result, the Euclidean time becomes 
$\tau\rightarrow \tau + \pi{\hat Q}_{\rm eff}$.
It implies that the radius of time like coordinate becomes small in the near horizon geometry. The solution  
(\ref{sl5}) describes a two dimensional laboratory black hole in the Plankian regime. 

\sp
Then the relevant two dimensional laboratory (Schwarzschild) black holes may be
obtained from the eq.(\ref{sl2}). The black hole solution is described by
\be
ds^2_L= dr^2 + \left (1- {{{\hat Q}^2_{\rm eff}}\over{r^2}} \right ) \left (1 + 
{{{\hat Q}^2_{\rm eff}}\over{r^2}} \right )^{-1} d\tau^2 \ .\label{sl6}
\ee
Since the black hole is obtained in the gravity decoupling limit $r\rightarrow {\hat Q}_{\rm eff}$, it describes the near
horizon geometry (\ref{sl5}). 
It suggests that the scattering of noncommutative strings \cite{kar-majumdar} 
may be seen to be associated with the production of two dimensional laboratory black holes (\ref{sl6}) at the
interaction vertex. These black holes are presumably short lived ultra-high energy states in the 
noncommutative string theory. They may may be seen to describe the $D$-string in the theory.

\sp
Then the Hawking temperature for the laboratory black hole (\ref{sl1}) is appropriately governed by the
the eqs.(\ref{hawk1}) and (\ref{hawk2}) in the gravity decoupling limit. It is given by
\be
T^{\rm Lab}_{\rm Hawking}={1\over{8\pi {\hat Q}_{\rm eff}}}\ .\label{hawk4}
\ee
It implies that the Hawking temperature increases as ${\hat Q}_{\rm eff}$ decreases by the emission of quantum
radiation from the laboratory Schwarzschild black hole geometry (\ref{sl1}). The process continues until all
the nonlinearity, $i.e.\ \Theta$-terms in ${\hat Q}_{\rm eff}$, are radiated out. Then ${\hat Q}_{\rm eff}$ 
takes a constant value ${\hat Q}$ and Hawking temperature attains its maximum $i.e.$ the Hagedorn temperature in the 
theory. 

%%%%%%%%%%%%%%%%%%%%%%%%%%%%
\section{Concluding remarks}
%%%%%%%%%%%%%%%%%%%%%%%%%%%%

To summarize, we have considered the evolution of gravity, starting from the classical to the Planckian regime 
on a $D_3$-brane in a string theory.
The noncommutative set-up being natural, it has incorporated a generalized notion to the various brane 
geometries. In particular the self-duality condition on the gauge field, 
in its counter-part with ordinary geometry, allows one to incorporate all order gauge curvatures in the theory. As a result,
the theory turns out to be an exact in $\Theta$. It has allowed us to explore some aspects of black hole geometries, 
including that of an exact conformal field theory obtained, otherwise, in an $SL(2,R)/U(1)$ gauged Wess-Zumino-Witten model 
\cite{witten91}. 
  
\sp
In context,  we have dealt with the problem in two steps. 
Firstly, we have exploited the existing noncommutative scaling in the cartesian coordinates of the frame-work,
to approximate the, otherwise, orthogonal spherical polar coordinates in the theory. 
It was shown that the radial coordinate $r$ and the Euclidean time 
$\tau$ receive lower bounds, of the order of Planck scale, in the theory.   
There, the spherical geometry, 
described by the angular coordinates ($\theta,\phi$), is rather modified to $S_2^{\Theta}$.
Interestingly, a transformation between a spherical polar coordinate ${\cal J}_S$-system and its 
noncommutative counter-part ${\tilde{\cal J}}_S$ was argued. It was shown that an orthogonal rotation of 
($r,\theta$)-plane, by an arbitrary angle $\theta$, in ${\cal J}_S$-coordinates leads to 
${\tilde{\cal J}}_S$-system. In other words, the broken spherical symmetry in the noncommutative set-up may be
restored by taking into account the complete configuration space of the rotation angle $\theta$. The underlying
idea was crucial to address a semi-classical solution in spherical polar coordinates in the noncommutative frame-work.

\sp
As a part of the second step, we have obtained a GS-black hole solution on a noncommutative $D_3$-brane in a 
static gauge. In the effective theory, the black hole was shown to be governed by a trivial noncommutative gauge 
field. On the other hand, a charged black hole underlying the notion of the GRN-black hole was obtained for a 
nontrivial, noncommutative gauge field in the frame-work. The black hole solutions were argued to be exact to all
orders in $\Theta$. Both, the GRN- and the GS-black holes contain higher order corrections, in $1/r$, with respect 
to their semi-classical cousins. Using a shift transformation, the obtained generalized solutions were re-expressed 
in the spherical polar coordinates. In the process, the mass and charge of the generalized black holes were shown 
to receive noncommutative corrections, which in turn may be viewed as a stretch in the event horizons of the GS- 
and GRN-black holes. Since $\Theta\neq 0$ even in the classical regime, the stretch in the event horizons was argued
to be a general phenomenon in the black hole physics. 
 
\sp 
Interestingly, different black hole geometries  were shown to be originated from the GRN-black hole for various values
of its effective mass and charge in the two extreme regimes.
For instance, the semi-classical regime in absence of an $U(1)$ gauge charge was shown to describe a typical 
Schwarzschild black hole. However, the presence of a gauge charge gave rise to
two independent semi-classical solutions. They are described (i) by switching-off the nonlinearity and (ii) by
switching-on the nonlinearity in the Maxwell theory. In absence of nonlinearity, the GRN-black hole in the
semi-classical regime was shown to govern a typical RN-black hole. However, the presence of nonlinearity 
in the ${\bf E}$-field, a priori, did not introduce any substantial change in the nature of the RN-black hole. 
It had modified the parameters in the RN-black hole geometry and subsequently introduced stretches in the event 
horizons. In addition to the stretch, the $\Theta$ parameter was argued to decouple the underlying $4$-dimensional 
effective space-time into two independent sectors, $i.e.\ L$- and $\perp$-spaces. Incorporating the appropriate
noncommutative scales for a fixed $\Theta$, the classical regime on the $D_3$-brane was shown to represent a 
two dimensional near horizon black hole geometry.  
At the other extreme, $i.e.$ in the Planckian regime, the GRN-black hole geometry was argued to be 
governed exclusively by the nonlinear ${\bf E}$-field in the theory. It was shown that the GRN-black hole
precisely reduces to the GS-geometry in the regime, which in turn gave rise to the two dimensional 
laboratory black holes in the theory. Our analysis suggests that the $D$-string in the Planckian regime
may be viewed as a two dimensional laboratory black hole. It is inspired by the one of old conjectures of
't Hooft in the context of a particle dynamics at Planck scale \cite{thooft}. 

\sp
In addition, we have revisited the phenomenon of Hawking radiation from the GRN-black hole
in the noncommutative formalism. Since different black holes were shown to be originated from the GRN-black hole,
the Hawking temperature in the theory may be argued to fluctuate from its ever increasing behaviour from the
classical to the Planckian regime. For instance, the temperature begins to increase with the typical 
Schwarzschild black hole, followed by a sharp fall to describe the  RN-geometry in the classical regime and 
finally it attains
the Hagedorn temperature in the Planckian regime. The passage from the semi-classical to the Planckian regime in the
noncommutative frame-work was argued to be governed by a series of quantum radiations due to the ${\bf E}$-field.
The increase in nonlinear ${\bf E}$-field, in each step, lead to the gradual decoupling of the (Schwarzschild) mass 
$M\rightarrow {\hat Q}/{\sqrt{G_N}}$, 
followed by a series of radiations in the Planckian regime. The decoupling of a series of
nonlinearity in the ${\bf E}$-field dominates at the Planck scale and finally the Hawking radiation ceases at 
${\bf E}_c$.

\sp
To conclude, the closed string decoupling limit at the Planck scale is a powerful technique to explain some of
the quantum gravity effects. Though Einstein's GTR is expected to break down at Planck scale, the string frame-work
appears to forbid the possibility. In other words, the GTR coupled to the nonlinear Maxwell's theory on a  $D_3$-brane
provides a plausible frame-work at Planck scale consistent with the special theory of relativity. 
At this point of time, perhaps the noncommutative frame-work provides an appropriate forum to address some of the 
open questions in quantum gravity.

\sp
\sp

\noindent
{\large\bf Acknowledgments}

\sp
S.K. acknowledges a partial support, under SERC fast track young scientist PSA-09/2002, from the
D.S.T, Govt.of India. The work of S.M. is partly supported by a C.S.I.R. research fellowship.

\vfil\eject

%**********************************************************%
\def\anp{Ann. of Phys.}
\def\cmp{Commun. Math. Phys.}
\def\prl{Phys. Rev. Lett.}
\def\prd#1{{Phys. Rev.} {\bf D#1}}
\def\jhep{J.High Energy Phys.}
\def\cqg#1{{Class. \& Quantum Grav.}}
\def\plb#1{{Phys. Lett.} {\bf B#1}}
\def\npb#1{{Nucl. Phys.} {\bf B#1}}
\def\mpl#1{{Mod. Phys. Lett} {\bf A#1}}
\def\ijmpa#1{{Int. J. Mod. Phys.} {\bf A#1}}
\def\ijmpd#1{{Int. J. Mod. Phys.} {\bf D#1}}
\def\rmp#1{{Rev. Mod. Phys.} {\bf 68#1}}

%**********************************************************%


\begin{thebibliography}{99}

\baselineskip= 18 truept

\bibitem{hawking05}S. W. Hawking, {\tt hep-th/0507171} (2005).

\bibitem{hawking1}S.W. Hawking, \cmp{\bf 43} (1975) 199.

\bibitem{gibbons-hawking}G.W. Gibbons and S.W. Hawking, \prd{\bf 15} (1977) 2752.

\bibitem{kraus-wilczek}P. Kraus and F. Wilczek, \mpl{\bf 9} (1994) 3713.

\bibitem{parikh-wilczek}M.K. Parikh and F. Wilczek, \prl{\bf 85} (2000) 5042.

\bibitem{polchinski}J. Polchinski, \prl{\bf 75} (1995) 4724.

\bibitem{gibbons-hashi}G.W. Gibbons and K. Hashimoto, \jhep{\bf 09} (2000) 013; 

\bibitem{gibbons-herdeiro}G.W. Gibbons and C.A.R. Herdeiro,  
\prd{\bf 63} (2001) 064006; 

\bibitem{gibbons-ishibashi}G.W. Gibbons and A. Ishibashi, {\cqg \bf 21} (2004) 2919.

\bibitem{mars-senovilla-vera} M. Mars, J.M.M. Senovilla and R. Vera, \prl{\bf 86} (2001) 4219.

\bibitem{kar-panda}S. Kar and S. Panda, \jhep{\bf 11} (2002) 052.

\bibitem{hashimoto-ho-wang}K. Hashimoto, P-M. Ho and J.E. Wang, \prl{\bf 90} (2003) 141601;
K. Hashimoto, P-M. Ho, S. Nagaoka and J. E. Wang, \prd{\bf 68} (2003) 026007.

\bibitem{kaloper}N. Kaloper, {\tt hep-th/0501028} (2005).

\bibitem{kar-majumdar}S. Kar and S. Majumdar, {\tt hep-th/0501067} (2005), to appear in \ijmpa.

\bibitem{nicolini}P. Nicolini, A. Smailagic and E. Spallucci, {\tt hep-th/0507226} (2005).

\bibitem{darabi}F. Darabi, {\tt hep-th/0508106} (2005). 

\bibitem{chu-lechtenfeld}C-S. Chu and O. Lechtenfeld, {\tt hep-th/0508005} (2005).

\bibitem{nasseri}F. Nasseri, {\tt hep-th/0508117} (2005); F. Nasseri and S.A. Alani, {\tt hep-th/0508117} (2005).

\bibitem{seiberg-witten}N. Seiberg and E. Witten, \jhep{\bf 09} (1999) 032.

\bibitem{minwalla}R. Gopakumar, S. Minwalla, N. Seiberg, A. Strominger, \jhep{\bf 0008} (2000) 008.

\bibitem{garcia}E. Ayon-Beato and A. Garcia, \plb{\bf 464} (1999) 25. 

\bibitem{tamaki}T. Tamaki and K. Maida, \prd{\bf 62} (2000) 084041;
H. Yajima and T. Tamaki, \prd{\bf 63} (2001) 064007.

\bibitem{kar-1}S. Kar, \npb{\bf 554} (1999) 163; \ijmpa{\bf 1} (2001) 41.

\bibitem{thooft}G. 't Hooft, \plb{\bf 198} (1987) 61; \npb{\bf 304} (1988) 867; 
\npb{\bf 335} (1990) 138.

\bibitem{ver-2}E. Verlinde and H. Verlinde, \npb{\bf 371} (1992) 246.

\bibitem{ver-21}E. Verlinde and H. Verlinde, {\tt hep-th/9302104}.

\bibitem{kar-maharana}S. Kar and J. Maharana, \ijmpa{\bf 10} (1995) 2733.

\bibitem{alvarez-gaume}L. Alvarez-Gaume and M.A. Vazquez-mozo, \npb{\bf 668} (2003) 293.

\bibitem{kmp}S. Kar, J. Maharana and S. Panda, \npb{\bf 465} (1996) 439.

\bibitem{eardley-giddings}D. M. Eardley and S.B. Giddings, \prd{\bf 66} (2002) 044011.

\bibitem{witten91}E. Witten, \prd{\bf 44} (1991) 314.

\bibitem{emparan}R. Emparan, \prd{\bf64} (2001) 024025.

\bibitem{emparan-horowitz-myers}R. Emparan, G.T. Horowitz and R.C. Myers, \prl{\bf 85} (2000) 499.

\bibitem{giddings-thomas}S.B. Giddings and S. Thomas, \prd{\bf 65} (2002) 056010.

\bibitem{kar-jain-panda}P. Jain, S. Kar and S. Panda, \ijmpd{\bf 12} (2003) 1593.

\end{thebibliography}
\end{document}